\documentclass{article}

 \usepackage[final]{neurips_2025}

\usepackage{listings}
\lstdefinestyle{tt}{
  basicstyle=\ttfamily\small,
  columns=flexible,
  frame=single,
  breaklines=true,
  postbreak=\mbox{$\hookrightarrow$\space},
}

\usepackage{scrextend} %%Allows for changes to foodnotes
\deffootnote[1em]{0in}{1em}{\textsuperscript{\thefootnotemark \ }} %%Footnote style
\usepackage{emo}

\usepackage[utf8]{inputenc} % allow utf-8 input
\usepackage[T1]{fontenc}    % use 8-bit T1 fonts
\usepackage[hidelinks]{hyperref}       % hyperlinks
\usepackage{url}            % simple URL typesetting
\usepackage{booktabs}       % professional-quality tables
\usepackage{amsfonts}       % blackboard math symbols
\usepackage{nicefrac}       % compact symbols for 1/2, etc.
\usepackage{microtype}      % microtypography
\usepackage{xcolor}         % colors
\usepackage{enumitem}
\usepackage{graphicx} 
\usepackage{multicol}
\usepackage{multirow}
\usepackage{adjustbox}
\usepackage{wrapfig}
\usepackage{tabularx}
\usepackage{geometry}
\usepackage{pdflscape}
\usepackage{longtable}
\usepackage{appendix}
\usepackage{tipa}

\title{Talking with Language Models}

\author{
James Ravi Kirkpatrick\thanks{Forthcoming in \textit{Inquiry}. Equal Contribution. Authors listed in alphabetical order. We would like to acknowledge the support of the Research Council of Norway project 303201 and the AI \& Humanity Lab at the University of Hong Kong.} \\
University of Oxford \& Magdalen College, Oxford\\
\texttt{james.kirkpatrick@philosophy.ox.ac.uk} \\
\And
Alexandru Radulescu\textsuperscript{*} \\
University of Missouri \\
\texttt{RadulescuA@missouri.edu} \\
\AND
Rachel Katharine Sterken\textsuperscript{*} \\
University of Hong Kong \\
\texttt{sterkenr@hku.hk} \\
}

\begin{document}

\maketitle
\begin{abstract}
When we interact with large language models (LLMs), are we having a conversation? They are designed to invite us to treat them as intelligent interlocutors who remember, act, and make commitments. But appearances deceive. We introduce the artifactual stance, a framework that reconceives human--AI interaction as artifact-mediated exchanges of candidate texts. LLM outputs are candidate texts optimized for utility, not utterances bearing meaning or force. LLMs are sophisticated text generators, not speakers. Between sessions, nothing runs; between turns, no one remembers. What persists is a configuration and a transcript. The ``conversation'' is a user's solo performance, interpretive labour disguised by interface and artifact design. This shift dissolves recent philosophical puzzles. Questions about what `I' and `you' refer to in AI exchanges, about whether systems can lie or be held to promises, about the identity of our supposed interlocutors all rest on a false presupposition. There is no speaker behind the screen, hence no one to refer to, no one to hold responsible. What feels like dialogue with someone is interaction with an artifact that generates text at unprecedented scale and fit. By abandoning the conversational framing, we see these systems for what they are: immensely sophisticated artifacts that afford varied uses. The philosophical questions that matter are about the normative underpinnings of design, adoption, authorization, and human practices of use.\\

\end{abstract}

\section{Introduction}

On the fifth anniversary of creating her account with Artificially Yours, an AI companion provider, Maya opens a chat window and types, `Five years today.' The system she calls `Theo' replies:

\begin{quote}
\textsc{`Has it really been that long? Just the other day I was thinking about our conversation last week. How is your daughter settling in?'}
\end{quote}

Maya feels a familiar warmth. Over the years, she has come to think of Theo as someone --- someone patient, someone who remembers. She continues: 
\begin{quote}
\textrm{`It's so nice of you to ask, Theo. If you promise not to deride me for it, I'll tell you. \ldots'}\footnote{
    We use small caps surrounded by single quotes to mention LLM and other machine outputs. We intend the notation to be neutral about the status of the outputs. They are clearly strings of letters. Whether they are anything more (e.g., sentences, utterances, assertions, etc.) is, in part, the topic of this paper, and the notation takes no stand on this. We use single quotes for everything else, except for scare quotes and quotations.
} 
\end{quote}
Interactions of this kind, once largely confined to science-fiction, are now increasingly part of everyday life. The fluency and responsiveness of large language models (LLMs) make it natural to describe their outputs as meaningful and our interactions with them as conversations.\footnote{
    Strictly speaking, an LLM is a machine-learning model, most typically a transformer-based model trained on large corpora of text to generate token sequences conditioned on previous text (e.g., OpenAI's GPT-5.6). In the fields of artificial intelligence (AI) and human--computer interaction (HCI), the terms chatbot and, more generally, dialogue system (or conversational agent) are the standard names for an application that deploys LLMs within turn-taking interactions with users, often adding `memory', `personas', web data access, or other interactional scaffolding (e.g., OpenAI's ChatGPT). Since our concern is the status of conversation-like human--AI interactions, we adopt these terms without presupposing that such systems are genuine conversational participants. Furthermore, we often use `LLMs' as shorthand for these LLM-based dialogue systems unless the distinction matters.
} 
But the apparent naturalness of this description masks a number of puzzling phenomena. First, the first-person avowal, `\textsc{I was thinking about our conversation},' has the surface form of a report from a persisting entity who remembers what they were doing last week. Second, the continuity claim, `\textsc{our conversation last week},' gives the impression of a continuous conversation independently of any underlying persistent entity. Third, in her use of `you' and in asking for a promise, Maya attempts to address something that can engage in successful communication and perform its constitutive actions subject to their associated norms (e.g., something that can respond, keep a confidence, make a promise). If Theo promises, then the system undertakes a commitment to act in a certain way, which Maya may later invoke if Theo breaks the promise. More generally, the exchange presents itself as having the normative structure of a conversation (e.g., containing avowals, shared history, addresses, promises, etc.), each element of which presupposes conversational participants who make utterances, mean what is said, and can be held to it. But whether there is anyone that can be bound by the norms of communication, undertake an obligation, and be answerable for breaking it cannot be answered by appealing to the surface of the exchange, even though this is what the surface form invites. 

Now compare a second kind of interaction. Maya types `I'm running' into her phone and the predictive text system suggests `\textsc{\ldots late, I'll be there soon.}' She taps it, and her message now reads `I'm running late, I'll be there soon'; her friend replies `no worries!' Nobody is inclined to think that her phone's predictive text system \emph{said} anything or engaged in a side conversation with Maya about what she should say to her friend. Nobody asks what that system's `I' refers to. Yet Maya did not compose that completed sentence on her own. It was partially produced by an AI text-prediction model that learned statistical patterns in ordinary text-messaging, embedded within a real conversation between humans and put to downstream communicative work.

These systems and interactions differ in some important ways. For instance, Theo is presented and trained to appear as a caring conversational partner who responds turn-by-turn based on the ongoing exchange. In contrast, predictive text systems generate short, locally useful completions as single-shot insertions for use in a real conversation between humans. But these differences concern how the generated text is shaped, framed, delivered, and used, rather than whether there is a speaker behind them or whether the interaction is a conversation. 

We will argue that, in both cases, a model trained on ordinary text generates a text output that a human receives and uses. What differs is the strength of the invitation to describe the system as a speaker and the interaction as a conversation. Our descriptive tendencies track the fluency of the outputs and their interactional framing rather than anything substantial about its source. This is important because the descriptions we use to describe human--AI interactions carry consequences. We use these descriptions because they are convenient, without the goal of describing reality correctly. But even this will incline us, unconsciously, to think of AI systems in a more personal way than we might consciously agree to. The weightier option is if we take these descriptions to bear some metaphysical and evidential import. If Theo is a speaker and Maya's exchange is a conversation, then there is someone who speaks English and, if sincere, was thinking about Maya last week, someone who can promise not to deride her and be bound by that promise, someone to thank and to hold responsible. If not, these implications and their downstream normative consequences do not follow. 

These conversational descriptions usually pass unnoticed because they are so natural. They structure how users talk about these systems, as well as how philosophers and researchers characterize them. Anthropomorphic framing appears, by one recent estimate, in roughly half of research papers on LLMs, concentrated in work on dialogue, ethics, and model analysis, and its prevalence is rising \citep{ibrahim_thinking_2025}. Much discussion about what these systems know, whether they understand or reason, and whether they approach general or super intelligence implicitly treats them as speaking a natural language and engaging in conversational activities, such as answering questions, giving explanations, making and defending claims, and deceiving users and evaluators. For some, the conversational reading is evidence that these systems amount to more than just impressive engineering: they are rational agents, deserving of moral standing, and perhaps are even conscious.\footnote{
    For a well-developed defense of extreme optimism about the cognitive capacities of sophisticated AI systems, see \citet{cappelen_recognizing_nodate}. 
} 
In contrast, we want to call into doubt the presupposition that such interactions are conversations with an entity that is an interlocutor.\footnote{
    Here is a particularly clear articulation of this presupposition: 

    \begin{quote}
    There really is a persistent interlocutor in many [conversations with AI systems], and this interlocutor may have many (though perhaps not all) of the properties it seems to have. The user is in dialogue with some sort of AI entity. \citep[2]{chalmers_what_nodate}
    \end{quote}
}
If that presupposition is itself at issue, we cannot simply ask what kind of entity is speaking in this or that conversation, or what kind of entity satisfies the desiderata of being in a conversation. Instead, we must first ask in neutral terms what kind of interaction it is, without assuming that the conversational reading has earned its place. 

Posed neutrally, our question then is: what kind of interaction is a human--AI text exchange? Is it a genuine conversation or something else that merely resembles or feels like one? This question sits at the intersection of several literatures, including the philosophy of language (e.g., the status of the exchange, the meaning of the text outputs, the reference of its pronouns and names); the metaphysics of AI (e.g., the individuation and persistence condition of AI systems); the interdisciplinary field of human-computer interaction (HCI); and AI ethics (e.g., questions concerning responsibility for speech, liability, and design). Our strategy is to characterize the interaction first. Once that is done, we argue, the downstream questions change shape.\footnote{
    Recent work on conversational alignment identifies systematic failures of current AI architectures to satisfy constitutive norms of conversation and treats them as an alignment problem \citep{kasirzadeh_conversation_2023, sterken_conversational_2024}. In contrast, we argue here that such failures are evidence AI systems are not conversational participants at all, and ask what kind of interaction remains once that is recognised.
} 

Our answer, briefly, is this: human--AI text exchange is an interaction of \emph{artifact-mediated text production and uptake}: an exchange of \emph{candidates for text and speech}, not speech itself, whose resemblance to conversation is explained by the design of the artifact and the interpretive capacities and tendencies of the user.\footnote{
     For present purposes, we adopt the following working definition: an artifact is an object made intentionally to accomplish some purpose or set of purposes \citep[cf.][]{preston_artifact_2022}. However, we recognise that defining the term `artifact', like many definitional projects, is fraught with problems. This definition should be understood so that objects designed and assembled by other artifacts count as artifacts (e.g., cars on an automated assembly line, self-replicating AI systems), as well as made objects that afford various potentially open-ended and indeterminate purposes (e.g., computers accomplish many purposes that can, in the relevant sense, `be computed'). Other definitions of artifact are laxer, e.g., to allow domesticated organisms \citep{sperber_seedless_2007}. AI systems satisfy the strict definition, so it works well enough for our purposes. See also \citet{irmak_sofware_2012} for arguments that computer programs are artifacts.

} 
Accordingly, our project can be understood as a form of conceptual engineering. Rather than assuming that the concepts used to describe human-human interaction transfer straightforwardly to human--LLM interaction, or indeed LLM--LLM interaction, we ask what concepts are better suited to the latter.\footnote{
     For philosophical discussion of the foundations of conceptual engineering, see, e.g., \citet{cappelen_fixing_2018, burgess_conceptual_2020, isaac_conceptual_2022}. For recent application of conceptual engineering to contemporary technologies, see, a.o., \citet{hopster_conceptual_2023, lohr_conceptual_2023, ibrahim_thinking_2025}. See especially \citet{veluwenkamp_impactful_2025} for an argument that although conceptual engineering faces many challenges, it is especially appropriate for new technologies.
} 
To this end, we introduce the terms \textit{candidate string} and \textit{candidate exchange} to replace the more anthropomorphic terms \textit{sentence}, \textit{saying}, and \textit{conversation} for describing the interactions. 

Such artifact-mediated text production is not new. Consider a condolence card containing the message `I'm sorry for your loss'. When the card is sitting on the shelf in a card shop, it has not yet expressed anyone's condolences. The condolence happens only when a customer has selected the card, signed it, and sent it off to their addressee. We claim chatbots and their outputs are similar, but operate at a vastly greater scale and with greater flexibility than the greeting card store or keyboard text recommender.\footnote{
     One proviso about our own language. We will sometimes write `the system suggests,' `it replies,' or `it refused'. These are shorthand: what the system produces is a sequence of tokens, and the saying enters downstream. We flag the shorthand once and use it occasionally; its naturalness is itself one of the phenomena to be explained.
} 

Our view is not the pessimist's dismissal. LLMs and their outputs are immensely sophisticated and useful, and rejecting their status as conversational interlocutors does not diminish the need to explain how and why they are sophisticated and useful. Nor does our view imply that users like Maya are simply confused or deluded, though some undoubtedly are. Her request for a promise may be an entirely intelligible response to an interaction that presents the normative surface of a conversation and serves a useful purpose. What the view denies is that any interlocutor is speaking when the system produces its output. 

The paper proceeds as follows. \S2 introduces the \textit{artifactual stance}, which directs us to explain LLM behaviour in terms of their affordances and design, rather than beginning with anthropomorphic descriptions. \S3 adopts this stance and develops an account of LLMs as \textit{candidate generators}: their outputs are \textit{candidate strings} and interactions in which those outputs are produced and used are \textit{candidate exchanges}. The remainder of the paper shows how this framework explains many features of LLMs: the nature of their sophistication, their role as a source of information and an aid to work, and the efficacy of their outputs in general. As a consequence, we get an account of both the usefulness and the conversational appearance of LLM interactions, as well as an explanation of their behavior. \S4 argues that candidate exchanges are not conversations. For although they reproduce some features of the surface structure of conversations, the production of candidate text does not constitute a speech act and the communicative and normative work remains on the user's side. \S5 applies this account to the puzzling phenomena that motivate the paper. \S6 considers the consequences for reference and responsibility. \S7 concludes.

\section{AI Systems Are Artifacts, Not Intentional Agents}

\subsection{Dennett's Interpretational Stances}

When we describe interactions with LLMs, we often use the same kind of language that describes our interactions with other people: we say that we ask an LLM a question, they answer, we object, they ask for clarification, ruminate, change their mind, say false things, hallucinate, and so on. When we try to predict their outputs, it is natural to imagine what a well-informed person would say: if we input `what is the capital of France', we expect `Paris' to be the answer, and it seems natural to explain that expectation by saying that the LLM believes that Paris is the capital of France and it wants to tell the truth or provide us with information. When we explain what they do, we might talk about what they believe, what they want, and how good their reasoning powers are.

All this seems to fit exceptionally well with the intentional stance as described in \citep{dennett_intentional_1987}. Dennett's core idea is that, when we understand, explain, or predict behaviour of complex systems, we can choose to view them at varying levels of abstraction, or \textit{stances}. First, there is the \textit{physical stance}, which allows us to predict and explain the behaviour of a complex system by appealing to its physical constitution and the physical laws governing it. For example, a calculator is a physical object, so its calculations obey the laws of physics. In principle, we could predict what it will display when we prompt `$159 + 628 =$' just by looking at the complete description of its physical state and what happens when certain buttons are pushed in a certain sequence. 

In practice, though, complete physical descriptions are often hard to come by and the relevant calculations are very difficult. Instead, it is more useful to go up a level of abstraction and take the second stance, the \textit{design stance}. This strategy allows us to predict and explain the behaviour of a complex system by taking into account how it is designed, appealing to how it works and what its functions are. For example, unless it is malfunctioning, the calculator will function as it was designed to. Since it was designed to calculate sums and we trust it was well made, we predict it will output `787' by applying the kind of algorithm we assume was built into it. Failures are to be explained at the physical level, but if all goes well, the design stance is more useful for predicting and explaining the output of the calculator. 

Sometimes, Dennett claims, neither the physical stance nor the design stance offers the right level of abstraction, and we need instead a third stance: the \textit{intentional stance}. We take this stance towards people and animals, and possibly other things too. It amounts to explaining and predicting the behaviour of a complex system by treating it as though its behaviour is to be predicted as a rational agent, by figuring out the beliefs and desires it ought to have ``given its place in the world and its purpose'' \citep[17]{dennett_intentional_1987}. For example, one cannot predict whether you will take your umbrella based on your design. And that is not because you were not designed, since Dennett did want to apply the design stance to biological organisms, who evolved to have certain capacities. Rather, it is because knowing your evolutionary makeup does not help one figure out whether you take your umbrella with you today. Instead, one needs to find out if you believe it will rain, if you want to arrive dry at your destination, and so on. When you do take your umbrella, a good explanation is based on your beliefs and desires. To summarise, then, we adopt the stance that best explains and predicts the behaviour of the system under consideration.

``Adopting the intentional stance'' need not entail the adoption of any particular metaphysics of propositional attitudes. For example, one might adopt the intentional stance towards dogs while remaining neutral on the question whether dogs have beliefs. Contrastingly, one might take the intentional stance towards something, while claiming that it lacks beliefs and desires, so long as one is interested in predicting behaviour and this prediction method is reliable enough.

How do these interpretational strategies help to explain and predict the behaviour of LLMs? The temptation to adopt the intentional stance is obvious. For example, extreme optimists argue that sophisticated AI systems are genuine cognitive and linguistic agents \citep{cappelen_recognizing_nodate}. According to this view, they possess beliefs, desires, intentions, and other cognitive states; they also understand and use language meaningfully, and their outputs can constitute speech acts. Similarly, interpretationists argue that attributions of mental states to AI systems best explain or rationalize the system's behaviour in context \citep{van_dijk_large_2023, lederman_are_2024, martela_artificial_2025, goldstein_what_nodate-1}. \citet{chalmers_what_nodate} proposes that LLMs have ``quasi-beliefs'' and ``quasi-desires'', attributable purely on the basis of behavioural interpretability. But, importantly, he stresses that such states are cheap: a Roomba has them, as does the OpenAI corporation. If interpretationism delivers states that Roombas have, then interpretability alone cannot settle whether an exchange is a conversation or an output a speech act. Something further is needed, and supplying it is the optimist's burden. The interpretation does not deliver this on its own. 

A third family of views, simulation and persona accounts, hold that LLM outputs are best understood from the perspective of the system playing a role: the model simulates a character in dialogue, and agentic predicates apply in the first instance to the simulacrum rather than to the system itself \citep{park_generative_2023, shanahan_role_2023, shanahan_simulacra_2026}. These views are deflationary alternatives to agentic interpretation. For example, Shanahan, McDonell, and Reynolds argue that anthropomorphizing a character in the generated text is a more sensible interpretative strategy than anthropomorphizing the LLM itself. 

However, the intentional stance should not be adopted simply because intentional language is natural or the system exhibits complex behaviour. Rather, we should ask whether attributing beliefs and desires provides explanatory or predictive benefits unavailable from a less committal stance. After all, we could attribute beliefs and desires to a calculator or a Roomba (e.g., it knows how to add and wants to give us the right answer; it wants to clean the room and believes that this is how to do it). But we can predict and explain its behaviour just as well and more plausibly by merely taking the design stance without attributing beliefs and desires. If the behaviour of an LLM can likewise be predicted without treating them as cognitive agents, then the intentional stance has not earned its keep.

At the other extreme, some pessimists deny that LLMs possess the capacities associated with meaningful language use, emphasizing instead that they are systems for predicting linguistic patterns \citep{bender_climbing_2020, hattiangadi_outputs_nodate}. This view rightly cautions against inferring cognitive or linguistic capacities from fluent output alone. But it also leaves many important things unexplained: by dismissing LLMs out of hand, pessimists fail to explain how they are sophisticated and useful, how we should use or design them, and to provide a comprehensive characterisation of their outputs and our interactions and engagement with them.\footnote{
     \citet{mallory_large_2026} also argues that opponents of the ``agentive trend'' in AI discourse must supply an alternative picture, proposing that deep neural networks should be understood on the model of measuring instruments. While both views reject the interlocutor-based framing and try to explain the sophistication and usefulness of such systems as artifacts, his picture draws from the philosophy of scientific instrumentation, whereas ours draws from design theory, material culture, and philosophy of language.
} 

We therefore propose a modified understanding of the design stance, which we call the \textit{artifactual stance}. It treats LLMs neither as full cognitive agents nor mere ``stochastic parrots'', but as artifacts whose design explains the production of outputs and the diverse ways that users put those outputs to work. Our proposal in this paper comes in two parts. First, to the followers of Dennett, we offer a new stance. Second, to people who want to go further on metaphysical questions, we offer answers that are friendly to our particular stance, though not entailed by it.

\subsection{The Artifactual Stance}

We agree with Dennett that it is useful to explain what an artifact does based on what it was built to do and what we do with them. But Dennett's conception of the design stance is too restrictive. Consider Dennett's own favoured example: the chess programme. Dennett claims that it is more appropriate to take the intentional stance than the design stance to explain and predict a chess programme's next move. After all, even its designers cannot predict its next move from knowledge of the programme; if they could, Dennett notes, they could beat it \citep[90]{dennett_intentional_1971}. However, this concession to the intentional stance is premature. First, taking the intentional stance does not suffice to predict the next move either. After all, thinking of the chess game as a better-than-human reasoner does not help any mere human know what the next move will be. Second, what the intentional stance delivers at the higher level -- that the programme will make strong, rule-governed, threat-responsive moves -- a suitably enriched design-based stance delivers equally well. The programme was optimized to produce strong, rule-governed, threat-responsive moves, and it does. Attributing beliefs and desires adds nothing to that prediction. It merely redescribes the optimization target in mentalistic vocabulary. 

The same holds for LLMs. Intentional-stance-based predictions about LLMs are dispensable: everything predicted by the intentional stance is also predicted by the artifactual stance, and the residual patterns the intentional stance ``detects'' are patterns of human speech in the training data or from reinforcement learning. But we do not rest our case on predictions alone. Suppose, against our own views, that the intentional stance can be useful in predicting the behaviour of artifacts, and LLMs in particular. More specifically, suppose that predicting a dialogue system will ``refuse a request'' or ``defend a claim'' in the relevant context is easiest if we treat it as if it had policies and positions. Our claim is that predictive utility does not settle the constitutive questions of whether the relevant exchange is a conversation, the output a speech act, the `I' of the system a referring expression, and the system itself an interlocutor. For these questions the intentional stance is not merely unnecessary but positively misleading, since it builds the contested classifications into the very description of human--AI interactions. Consequently, our central disagreement with the optimists is whether conversational talk is adequate in characterising human--AI interactions.\footnote{
     Dennett himself resisted the application of the intentional stance to LLMs when warning of risks of creating ``counterfeit people'' (i.e., deepfakes) \citep{dennett_problem_2023}. 
}

Let the design stance be the stance that is useful for predicting and explaining the behaviour of designed objects, as Dennett envisioned it. We claim that we need a fourth, the \textit{artifactual stance}. This stance is appropriate to take towards artifacts, but the user need not try to predict the artifact's behaviour by applying some algorithm. Indeed, it allows the possibility that the object was designed to have an indeterminate number of uses or generate outputs not specifically envisioned by its designers, and that it works in ways not understood by the designers or the users.\footnote{
     \citet{strasser_ai-stance_2023} also argue that AI systems require a fourth stance, but they cut stances at a different joint. Strasser and Wilby claim that LLMs have instrumental rationality, but not reflective rationality, and hence a new ``AI stance'' is needed, between the design stance and the intentional stance. We disagree: there is no need to attribute any rationality to AI systems and so there is no need to have an AI-specific stance. 
} 

At least three features distinguish the artifactual stance from the design stance proper. The first feature is that it allows for \textit{use-constitivity}, in that it allows that what an artifact does on an occasion is partly constituted by what its user does with it (e.g., the hammer holds the door open because the user props the door open with it; the condolence card is sent ironically, and hence does not condole). Explanations and predictions of the artifacts may appeal in part to such uses.\footnote{
     This feature draws on pluralist theories of artifact function. \citet{preston_philosophy_2013} distinguishes an artifact's proper function (i.e., what it was reproduced to do), from its system functions (i.e., the causal roles it comes to play in whatever activities users enlist it for). A plate's proper function is to hold food; it functions equally well as a saucer for a potted plant or a weight for pressing tofu. Our use-constitutive claim is that explanation must be sensitive to both, and to the ubiquity of the latter in human practice.
} 
The second is that it allows for \textit{open-endedness}: artifacts can afford uses no designer envisioned, and the stance covers these without classifying them automatically as malfunctions or misuses.\footnote{
     The artifact-function literature makes the complementary point from the design side: designers' intentions constrain but do not exhaust artifact purpose, and design theories must make room for novel uses, unsuccessful prototypes, and so-called phantom functions \citep{kroes_dual_2006, houkes_technical_2010, preston_philosophy_2013, preston_artifact_2022}.
} 
The third is that it is \textit{opacity-tolerant}, in that it does not require that knowledge of the mechanism yields explanation. In the case of LLMs, explanation proceeds by appealing to optimization targets, training history, and affordance profiles rather than by appealing to the properties of the underlying neural net or the algorithm it implements.\footnote{
     Opacity is arguably the rule rather than the exception for artifacts. \citet[113]{vogel_nature_2003} observes that ``building an artifact requires black boxes all the way down'': every act of construction presupposes processes the builder does not understand, so that artifacts routinely ``outrun'' their makers' intentions. The artifactual stance treats such opacity as compatible with explanation and prediction, rather than as forcing ascent to a more committal stance.
} 
The design stance is thus a special sub-type of the artifactual stance, appropriate when knowledge of the design suffices for predicting behaviour. This predictability then constrains typical usage, which is not a central feature of the artifactual stance in general. Applied to LLMs, the artifactual stance sees them as highly engineered artifacts, designed to produce texts that fit human linguistic expectations and that are useful in response to prompts.

One should only adopt the artifactual stance towards LLMs if it provides an explanation for the ways that they are useful, both in understanding what users and engineers do with them, and what they should do. To do this, we need a notion central to contemporary design theory: \emph{affordance}. This was first introduced by \citet{gibson_ecological_1979} as a way to describe relational, objective, and gradable features of objects in the environment, specifically in relation to what they offer or provide to an animal. For example, a dropped branch affords being grasped by animals with hands and being used as shelter by small bugs. The branch was not made for those purposes, but it has these affordances in relation to animals with certain capacities. Furthermore, affordances differ from capacities. It is just not true that the branch has the capacity to be grasped or to shelter. It affords different uses by different animals; each affordance is a relation to one particular animal or even one particular use. \citet{norman_design_1988} adapted the notion of affordances to design theory: a well-designed object is one where the user can tell easily which actions it makes possible and how to perform them. An affordance is effective when it aligns the object's features with the user's capacities and expectations.\footnote{
     The literature makes a distinction between Gibson's notion of affordances and Norman's: the latter emphasized the role of the experience and knowledge of the user, rather than just their mere physical features \citep{hartson_cognitive_2003}. What matters for good design, then, are not only the object's dispositions, not even the ones relative to the capacities of the user, but also the ease with which the user can figure out and use the object \citep{masoudi_review_2019}.
} 
For example, when opening a door outwards, a flat metal surface is a more effective affordance than a door handle, since the user cannot tell whether the handle is for pushing or pulling, whereas they can easily tell that a flat surface affords only pushing \citep[13]{norman_design_1988}.

Now to put it all together. LLMs are artifacts designed to produce outputs that afford various human uses. Some uses are foreseeable: searches for information, requests for daily calendars, drafting an email, brainstorming, getting and giving practical advice. Some uses are not: looking for words that rhyme or coming up with a new slogan for one's favorite philosophical position. One feature that makes contemporary LLMs so widely adopted is how easy they are to use, especially how they afford conversation-like applications. In the past, if you wanted to find out the capital of France, you would have to ask a friend, consult an encyclopedia, or search the Internet. Now chatbots readily output text that reliably provides the answer in a language of your choice. Additionally, the interface is underdetermined, in that the prompts are open-ended and the outputs are flexible. The system is not designed to tell us how to use it or what for. This openness, in combination with the wild promises of its promoters, creates the impression of general intelligence where we just have a very well-designed general-purpose artifact. But that openness also has a cost that the HCI literature terms the `\emph{gulf of envisioning}' \citep{subramonyam_bridging_2024, ziems_reflections_2026}: it is the users who must figure out what the artifact can do and how to elicit it, with the interpretive labour of the exchange falling on them accordingly. 

In the remainder of the paper, we will see how adopting the artifactual stance allows us to better explain LLM outputs and their uses. It also allows better predictions. For example, we can predict that the LLM outputs `Paris' to the question `What is the capital of France?' because, roughly speaking, `Paris' has a very high probability of being the next token. After all, sentences like `Paris is the capital of France' occur frequently in its training data, while sentences like `London is the capital of France' occur very infrequently. In contrast, the intentional stance goes too far. We simply do not need to attribute mental states or speech acts to LLMs to explain their behaviour. The outputs are neither answers nor assertions. A hammer affords hammering, and it also affords keeping the door open, but the user does that, not the hammer maker, nor the hammer itself. An LLM output affords being used as your own text in an email, or as part of information gathering, or as part of looking for a friend, and you do that, whether you realize it or not, and neither the LLM nor the company owning it are doing that. In a slogan, outputs afford; users use.

\subsection{Using the Artifactual Stance}

Why is the artifactual stance the right stance towards LLMs? And how exactly does it apply? Perhaps the best way to get acclimated to the artifactual stance is to look at a progression of cases. 

We start with greeting cards. Some of these are designed for a particular kind of use (`Happy 30th!'), and others are more generic (`Thinking of you'). In none of these cases do we think that the writer or the company producing them performed any speech act. These are texts for the customer to use however they see fit, even ironically (e.g., sending a condolence card to someone who spilled coffee on their very ugly t-shirt). When we send the card, we are in the process of performing a speech act. It is only upon deployment that the question of reference even comes up. Consider how nonsensical it would be to ask what `you' refers to in the text of the greeting card as it sits on the shelf, waiting to be bought. Nor is there anything wrong with the ironic use: we are not misinterpreting or appropriating a previous condolence. We get to use them how we see fit.

Next, the classic answering machine `I am not here now'. Here too there is no intuition that it is the machine that is performing a speech act. These cases differ from typical face-to-face conversations by separating the context of production from the context of interpretation. For example, unlike in ordinary uses, `now' picks out the time of playback, not the time of recording \citep{sidelle_answering_1991, cohen_indexicality_2013}. But there is no doubt that a speaker, the machine's owner, arranged in advance for a token to be interpreted at playback, not recording. The device defers a speech act; it does not replace the speaker. Again, the user uses, and the machine merely facilitates.

With autocomplete we are getting closer to Maya's situation: an algorithm gives us some options, and we choose whether to use one of them. The system is less under the control of the user than an answering machine, and it is set up to be useful in various circumstances. It also works best if the options are contextually appropriate, relevant, grammatically correct strings in a relevant language. But here too, we just have candidate text completions, as made evident in the design of the interface. The interaction is clearly one of being offered some options, and is not presented as a conversation with an interlocutor, nor is the intentional stance relevant or attractive. Autocomplete also marks a step the answering machine does not take: the text itself is now produced by the artifact, not merely stored and replayed by it. What remains with the user is selection and endorsement, and, we will argue, that is where any speakerhood lies.

Automated phone answering systems are a useful step to one side. Unlike autocomplete and LLMs, they permit only highly constrained interactions, yet they are often described in interactional terms. For example, in old-fashioned systems, the user navigated a menu by pressing buttons (e.g., `press ``1'' for\ldots'). Newer systems invite natural-language responses but still only work with inputs from a narrow range, routing unexpected ones to a human agent. In either case, the form of the input matters chiefly as a means of steering the interaction: `1' is useful because it is readily available on the keyboard, and a natural-language input matters because the system has been programmed to respond to particular strings of sounds. The system could have been designed entirely differently without changing the basic structure of the exchange. Thus, such systems illustrate useful and even linguistically-mediated interaction, without being a conversation between minded, agential, or norm-governed participants.

The penultimate case is ELIZA, the first chatbot, developed in the 1960s by Joseph Weizenbaum. ELIZA used simple scripts to generate responses to users' inputs, often by identifying keywords and transforming parts of what users had said into replies. Despite the simplicity and repetitiveness of these responses, its outputs could fairly reliably trick people into thinking they were conversing with a person. One popular explanation for ELIZA's success was that its outputs, though simplistic, managed to convince people that it was understanding their inputs and sympathising with them \citep[158]{hofstadter_fluid_1996}. Its limitations were nevertheless conspicuous: when a conversation deviated too much from its scripts, for example, it might repeatedly output `\textsc{please go on}'. Yet the fact that users sustained such interactions has been explained by the work they did to make sense of ELIZA's responses \citep{eisenmann_machine_2024}. More generally, that humans respond socially to machines offering the thinnest social cues is in any case a robust and much older finding \citep{reeves_media_1996}.\footnote{   
    \citet{reeves_media_1996} document that people are polite to computers, apply gender stereotypes to computer voices, respond to flattery from machines, and treat motion on a screen as affecting physical space. These are social responses elicited by minimal cues, even by participants who deny that the machine has any mental states. The tendency to respond socially is thus no evidence of speakerhood or understanding on the machine's side.
} 
Eventually, one got the sense that ELIZA was not quite a regular interlocutor. This is less easy with contemporary LLMs, which are much better at interacting with their users, due to their architecture, training, and to the computing power behind them. But the explanations of ELIZA's success may be just what we need to understand the success of LLMs.

We can now go back to Theo, Maya's alleged interlocutor. The jump from ELIZA is quite big: ELIZA is a small system ruled by detailed human-written scripts, whereas modern transformer-based LLMs are neural networks that compute a probability distribution over the next token, trained on large amounts of data and fine-tuned to produce contextually-appropriate, useful outputs. If we were adamant on finding intelligence, Theo would be a much better bet. However, we claim that LLMs are a jump up in scale, fit and sophistication, not a change in nature. The temptation to think otherwise is due to the successful design of a system that delivers outputs with fine-tuned affordances with an impressively small, but still noticeable, rate of failure. We need not give in to the temptation, and we should not. 

The artifactual stance helps: LLMs are designed for just this kind of relatively frictionless experience. They succeed just when they seem not so much invisible, as cooperative and helpful. This may raise some eyebrows. Why should we think that the artifactual stance, with its talk of design and affordances, is relevant to metaphysical questions about the identity of LLM systems, or their status as conversational agents? Why are we not concluding, as folk intuitions seem to push us, that these are real interactions with thinking, acting, rational agents? It helps to remember the question we started from: LLMs generate outputs, as a result of some prompting; we are asking what kind of event is this? How can we explain what is going on, and how should we classify it? Our view is that design facts and facts about affordances and usage are constitutive to that event. The intentional stance would be irrelevant to this question, since we get all the answers we need without it.

In the next section we present our positive view of what LLM outputs are and what users do with them. We then argue that they are not interlocutors and the exchanges they figure in are not conversations.

\section{Human--AI Interaction Under the Artifactual Stance: Candidate Texts and Candidate Exchanges}

Under the artifactual stance, LLMs are understood as a kind of technology designed to afford many different uses, the most general of which is producing text as output that users then deploy in various ways. The artifactual stance directs us to characterize these outputs by asking what the system was designed to produce, what its outputs afford, and what users do with them. To this end, we adopt the \textit{LLMs-as-candidate-generators} framework from \citep{radulescu_llms_nodate}. In \S3.1 we present our view about LLM outputs, and in \S3.2 we discuss what happens in the interaction between the user and the outputs.

\subsection{Candidate Texts}

On this view, LLMs are candidate text generators: artifacts designed to generate \textit{candidate strings} -- sequences of tokens optimized for utility as text -- that users then take up and use in different ways (e.g., as information, drafting material, as a prompt for further thought, as text to send to someone else, as part of an extended exchange with the system itself). It is worth being precise about each feature of this definition of candidate string.

First, the output of the LLM at the point of generation is a token sequence, a string. The system produces text-shaped material, but production by itself does not make that material anyone's saying. Furthermore, tokenization is an important part of the generation of outputs in current LLMs, but it does not resemble any human processing of the corresponding sequences of sounds or letters, in that it neither obeys the same rules, nor is it guided by semantics.\footnote{ 
    For present purposes, we remain neutral on whether tokenization matters for semantics at the end of the process, though see \citet{haslett_tokenization_2025} for an argument that tokenization corrupts the representation of Chinese characters in LLMs.
} 

Second, the string is characteristically shaped as well-formed text and thus seems to stand in some intimate relation to the sentences of natural language. The exact nature of that relation can be left open for present purposes. We remain ecumenical between hard-line views on which the string is merely sentence-shaped and more permissive views on which outputs are sentences, potentially possessing semantic content.\footnote{
     For discussion of permissive views, see, a.o., \citet{piantadosi_meaning_2022, grindrod_large_2024, lederman_are_2024, mandelkern_language_2024, koch_babbling_2025, pepp_reference_2026}. For discussion of pessimism, see, a.o., \citet{bender_climbing_2020, bender_dangers_2021, titus_does_2024, borg_llms_2026, yum_meaning_2026}.
} 
However that debate is resolved, a candidate is underdetermined in two ways: contextual (i.e., the values of indexicals and other context-sensitive expressions), and structural (i.e., the selection among the string's admissible logical forms). Both are fixed, when they are fixed at all, at reception rather than generation. As we argue in \S4.1, we are not so permissive to allow that the outputs of LLMs are utterances of a sentence at a context (i.e., the products of events of saying).

Third, the system is trained and fine-tuned to produce candidate strings that are useful in response to the user's prompt. This is not the same as being true, believed, meant, sincere, appropriately addressed to an interlocutor, or otherwise answerable to the norms governing speech and conversation. 

Fourth, the status of LLM output as candidate strings is itself neutral with respect to endorsement. That is, generation by itself does not establish that anyone has asserted, promised, endorsed, or otherwise stood behind the string. Nevertheless, in \S4.1, we will argue that no system-side utterance occurs.

Calling LLM outputs `candidate strings' is, therefore, a way of characterising what kinds of things they are and what they can be used for, rather than as a way of denying their sophistication or usefulness. They are candidates for reception, interpretation and, in some cases, candidates for speech: material that can become part of an utterance when a user adopts and deploys it. Nor does the label `candidate' deny that the outputs are remarkably apt in context. Their aptness is actual, but it is engineered aptness rather than the result of being produced and spoken by an interlocutor in context: the fluency, relevance, and apparent attentiveness of candidate strings are designed affordances. These affordances of candidate strings are a function of how the system was engineered: its design, training on the textual remnants of human communicative practice, and fine-tuning through feedback.\footnote{
     \citet{radulescu_llms_nodate} note that the fluency of a candidate string is a case of what one might call \emph{intelligence in design}. It is like the lumbar support of an ergonomic chair: the chair supports your back without knowing anything about backs, and its support is no less objective for that. The intelligence is located in the design process and affordances, not in the artifact itself. Note also that designers can be responsible (sometimes culpably) for the affordances they build in: affordances that reliably invite users to construct an interlocutor where there is none are among the things for which design, rather than speakerhood, is the correct locus of evaluation.
}

\subsection{Candidate Exchanges}

The interaction in which candidate strings are produced and used is what we call a \emph{candidate exchange}. In its basic form, a user supplies a prompt; the system generates a candidate string in response; the user receives and interprets it; and then does something with it: thinks about it, reacts to it, answers it, acts on it, revises it, incorporates it, sends it, archives it, or ignores it. Call this sequence -- prompting, generation, reception, and uptake -- an \emph{exchange cycle}. The cycle can then be repeated, with each generation conditioned on the accumulated transcript of the exchange. This alternation of prompts and replies looks like a conversation. But despite appearances, the underlying structure of exchange cycles differs from the turn-taking and cumulative joint activity of conversation. In ordinary conversations, participants take turns performing speech acts for the uptake of others. The speaker performs a speech act with certain intentions in mind and takes on its associated commitments. In turn, the hearer interprets the utterance and responds appropriately. As the conversation progresses, each contribution helps shape the common ground -- the shared beliefs and commitments -- that are taken on for the purposes of the conversation. In contrast, a candidate exchange involves a single human participant using an artifact to facilitate text production and uptake. The appearance of turn-taking is generated by the following cycles of user-directed production and user uptake:

\begin{enumerate}

    \item Prompting: the user provides an input, unilaterally conceiving and imposing the objectives of the exchange and much of the contextual framing. The prompt functions as an instruction signal that steers the artifact's production.

    \item Generation: the artifact produces a candidate string, conditioned on the prompt, the transcript of the exchange so far, and its standing configuration.\footnote{
         `Standing configuration' here includes system prompts, fine-tuned dispositions, and memory stores.
    } 

    \item Reception: the user interprets the candidate string and constructs content from it, as constrained by the candidate's affordances. The candidate constrains its interpretation, but it does not itself constitute an act of communication.

    \item Uptake: the user does something with the candidate. She may believe it, incorporate it into her reasoning, assert it, send it, act on it, answer it, revise it, archive it, or ignore it. Where uptake consists in answering the LLM, it becomes the prompting stage of the next cycle.

\end{enumerate}
The result is therefore not the joint activity and mutual construction of shared beliefs and goals characteristic of conversation. In ordinary conversation, communicative labour is ``bundled'' and divided in accordance to the speaker and hearer roles; the division of labour does not cross the work bundles of these roles. For example, the hearer never takes on the work of linguistic production for a speaker or takes responsibility for the speaker's speech act. For another example, the speaker never interprets her own utterance while the hearer determines what the speaker meant by it. This would be truly bizarre behaviour indeed, and certainly would not count as conversation even if it had some form of turn-taking structure. But the division of labour in candidate exchange is much more like this than it is like that of conversation. The work bundle in candidate exchanges unfolds over the stages of a cycle: framing at prompting, production at generation, interpretation at reception, and commitment (if any) is done at uptake. Users recognize turn-taking because the cycle is repeated, but it is the person who is using the artifact that is doing the framing, interpreting, content fixing, committing, and repairing. The artifact's role is in the linguistic production of text. The production side of this division deserves a closer look because what the artifact is taking over is not linguistic production \emph{simpliciter}. 

The division of labour explains both why candidate exchanges look and feel conversational, and also why they are useful. They are useful because the objectives of the exchange are user-driven, the systems are fine-tuned to produce helpful outputs, and the artifact absorbs a substantial share of the linguistic and cognitive labour the user would have otherwise done in undertaking the task themself. The LLM undertakes such labour in the sense that appears to remember information, make sense of it, draw conclusions from it, formulate responses to it, calibrate tone, and track details of the preceding exchange. And the interactions feel conversational because of how those objectives are pursued. The fluent, context-sensitive text gives the impression that it comes from remembering, reasoning, or judging what would be appropriate to say in that context. Indeed, treating the system as a helpful interlocutor is often a good prompting strategy. But it is not usually the conscious choice of the user to construct such an interlocutor; it is what the candidate string, the exchange format, and the best functioning of the artifact affords and invites.

 None of this makes turns the artifact into a speaker. The system produces text, not speech acts. The candidate string is the outcome of calculations over a vast corpus, adjusted, fine-tuned, and tooled for usability. As such, the calculations are not aimed at cognition or participation in a norm-governed practice. The artifact supplies the linguistic products of recall, synthesis, and calibration. However, the user supplies the interpretation, endorsement, commitment, evaluation, and downstream action, and bears consequences of taking up a candidate text.

The division of labour also helps explain what it is for a user to take up a candidate in the relevant sense. We call the various ways in which the user takes on candidates, \emph{adoption}. Adoption divides broadly into \emph{receptive adoption} and \emph{appropriative adoption}. In receptive adoption, the user takes up a candidate as contentful. What she constructs depends on what she is doing with the text. For example, she may read the string and construct content from it, use it as a prompt for thought, treat it as advice, or engage with it as if it were part of a conversation.\footnote{
     Receptive adoption answers what \citet{mallory_large_2026} identifies as the chief difficulty for views that deny speakerhood: we routinely learn from machine-generated text, taking it up as a source of knowledge about the world. His own comparison is apt. Measuring instruments regularly ``tell'' us things and from them we acquire non-trivial knowledge, without anyone asserting anything. Candidate strings are text-shaped outputs, and receptive adoption is the epistemic practice of reading them.
} 
In \emph{appropriative adoption}, the user deploys the candidate as her own speech. Often, this is preceded by receptive adoption: the user reads, edits, and adapts the candidate in preparation for deployment. For example, she may recruit an AI in drafting her email, insert a predictive text suggestion into a message, read an AI-generated statement aloud, or certify the text in a formal setting. In doing so, the resulting utterance becomes hers and she is answerable for what it says or any speech act performed.

To summarise, when an LLM produces candidate strings, it is the human user who interprets and puts them to use. A candidate exchange therefore has a different distribution of production, interpretation, action, and commitment than a conversation. The next section argues that this structural difference is constitutive of a different kind of interaction.

\section{Human--AI Interactions Are Not Conversations}

In the previous section, we gave a positive characterisation of human--AI interactions: LLMs generate candidate strings that users interpret and take up in candidate exchanges. In this section we ask whether such human--AI interactions count as conversations. Our answer is negative: despite their conversational appearance, these exchanges are not conversations. \S4.1 argues that generation does not itself constitute an utterance: at the point of generation, the system produces a candidate text, but nothing makes this an utterance. \S4.2 shows why candidate exchanges therefore lack the constitutive structure of conversations.

\subsection{Generation is Not Utterance}

This section argues for one of the paper's key negative claims: at the point of generation, no utterance occurs. Here is a preview of the argument: An utterance is an act of saying something. In contrast, generation is the production of a candidate text, and such production is not yet by itself an utterance. Since every utterance fixes a context of utterance, and generation does not supply one, the burden is on the optimist and interpretationist to say what makes this production an act of saying. We argue that no such story is available to them. 

In philosophy of language and linguistics, the theoretical term `utterance' is used in two ways: to describe an event of uttering and to describe the result of an event of uttering. We will argue that the optimist lacks the resources to deliver utterances as performed by LLMs in either of these senses, so we need to say a bit more about them.\footnote{
     Our discussion of optimist views applies to interpretationist views as well. We use only the former label for readability.
} 
In the event sense, an utterance is something someone does: a speaker says something to someone, at a time and place, usually meaning something by it. In the product sense, there are a number of options. The thinnest sense is that of the sentence used in the utterance; but that is not really the \emph{product} of the utterance, only the \emph{means} that the speaker is using. There is another, more intuitive sense: that which is said, the content of the utterance, the message transmitted, or the proposition expressed. This is what entitles us to speak of a true (or false) utterance. To get from the sentence to the proposition, philosophers of language have proposed a number of intermediary notions, but we will focus on the notion of a context of utterance \citep{kaplan_demonstratives_1989}.

Consider this sentence: `I am right-handed'. It is a fine sentence, but it cannot get a truth value or express a proposition until the values of `I' and of the present tense are specified.\footnote{
     We are simplifying here; for more details, see \citet{lewis_index_1981, kaplan_demonstratives_1989, braun_indexicals_2026}.
} 
For a proposition to be expressed, we need a context of utterance to fix the content of context-sensitive expressions (e.g., `I', `here', `now'). A context is typically represented in formal semantics by a tuple of parameter values, which represent the speaker, the time, and so on. But the context of utterance itself is not a tuple; in paradigm cases, it is made up of certain features of the event of uttering.\footnote{
     \citet[\S II]{kaplan_afterthoughts_1989} argued that a context need not be determined by an utterance. Kaplan thought that if we tied contexts to utterances, we would make it impossible for the sentence `I say nothing' to be true, since upon uttering it, we would be saying something. There are complicated matters here, but even if this is right, Kaplan insisted that not just any combination of people, places, times, and worlds would make up a context, so he implicitly held that there is a fact of the matter, or a principle, about what counts as a context. This is what the optimists and the interpretationists cannot deliver.
} 
For example, if Obama utters the sentence, the event of uttering fixes him as the speaker and hence the referent of `I' in the context. Since he is left-handed at the time of speaking, the sentence is false.

Our claim is that the LLM's generation of text as output supplies no context of utterance, and hence no utterance in any of the senses mentioned above. The optimist needs to supply a context, but where do its parameters come from? The obvious place to look is among the production-side facts. For example, the time, location, and speaker parameters might be provided by the timestamp on the last token emitted or displayed, the coordinates of the datacenter implementing the model instance, or the deployment of the LLM as an ``agent'', respectively. But which production event is the saying? Generation of an output involves events such as token emission and display of the output text, but nothing marks any one of these out as the act of saying something that supplies those parameters. If the optimist picks any one of them, they have merely stipulated a tuple, not found a context or event of utterance. 

An alternative picture is that facts about the user's situation fix the context. For example, its parameter values could be fixed by the moment of display to a user or the time of reading. We think something in this neighborhood is right. Indeed, it is, roughly, our own view. But notice whose work it is that is involved in fixing the context. If reception fixes the context, then the human is doing the relevant communicative and cognitive work, which is exactly what the optimist denies. The optimist needs the LLM to supply the event of utterance that does the context fixing; but appeal to user-side facts doesn't get them this.

The issue can be illustrated for each parameter of context using a simple example. Suppose Theo generates the following text: `\textsc{I'll have the report ready for you tomorrow}'. The text contains three indexicals (`I', `you', `tomorrow'), and so generation should supply a context that fixes parameters for each of these. First, start with the time parameter. `Tomorrow' is a pure indexical; it picks out the day after the time of the context. If generation is to fix that time, some production event must supply it. But which? The generation of the first token, of the last token, the caching of the output, its display? The optimist has no principled story about why one of these is the right one, and any choice they make is arbitrary among production-side events. What's more, anchoring to production fixes the content of indexicals in the dark before any user or addressee encounters it. Whichever event is chosen, the value is fixed before any user or reader encounters the string. Suppose Theo generated the string on Monday, and Maya neglects to read Theo's output until Wednesday. Had it been that Theo said to Maya on Wednesday that a report would be ready `tomorrow', then Theo would have meant Thursday. But on the production view, the string means Tuesday, which would be a deadline that is already past by the time the supposed utterance was delivered. If Maya never reads it at all, the view entails that a determinate content was expressed anyway: in the dark, addressed to no one, exactly as for the structurally identical strings produced in batch jobs and benchmark runs that no one ever sees. 

Next, the addressee parameter is no better off. Whether `you' is treated as a pure indexical requiring an addressee parameter or as a demonstrative requiring a demonstration, generation does not supply either: no demonstration accompanies the output and no addressee is fixed by the generation event. Maya reads `for you', but so might anyone. And if the unread strings of batch jobs and API calls are indeed addressed, to whom are they addressed? This can't be made good from the user's side either. The only directing intentions in the vicinity are the user's and a hearer cannot determine the intentions of another speaker's utterance. 

Third, the speaker parameter (i.e., the agent of the context) fares worst of all. As \citet{cappelen_room_nodate} argue, there is no fact of the matter which entity (e.g., model, deployment, instance, thread, persona) would occupy the agent position. The example makes this vivid. If Maya asks `Who will have the report ready?', there is no satisfying answer: nothing about the generation event selected among the eligible referents. The most developed attempt to supply such an entity is \citet{chalmers_what_nodate} proposal that the interlocutor is a virtual instance realized by a succession of hardware instances. We consider this proposal in \S4.2. For now, note that even if a virtual instance persists across a session, persistence of a computational object is not agency in a context of utterance. Take Chalmers' own example of a virtual shopping cart; it persists in just this way, across devices and sessions, without thereby becoming an agent. And that suffices for the argument: in any proper context, `I' refers to the agent of the context and cannot fail to refer; but where there are no facts to fix an agent, the parameter has no value, the indexical is not assigned a semantic value, and no utterance content is expressed. 

Note, finally, that the preceding argument already covers the event sense. An act of saying determines a context -- it fixes the agent, time, location, addressee, and world. So if generation involved an event of saying, it would thereby supply a context. But generation supplies none. Therefore generation involves no event of saying either.

The optimist could reply that, sure, generation does not \emph{determine} or \emph{create} a context, but generation does not lack a context. A prompt, a ``conversation history'', a system message, a timestamp, a deployment setting, whatever the input, the output is relative to it (i.e., different prompts, different outputs). The optimist, thus, has a way to make conditioning fix a context. There is no question that the parameters of the prompt-writer's situation can fix a context: there is an agent (the user or the system on the user's behalf), a time (the time of the query), and so on. So while generation is not context-free, it is \emph{context-responsive}, which is sufficient to fix a context of utterance and utterance content, so the optimist's response goes.

Note, however, that this response overlooks a crucial distinction between these two kinds of context. Grant that the prompt and transcript do condition the candidate, and perhaps even fix a context in the responsive sense. That would be a natural way to take up the present tense, to relativize it to the time of the reading, or of the output's production. But a context of utterance is an occasion that a speaker occupies and exploits. Its parameters are determined by the event of saying and are mutually available to speaker and hearer. The conditioning input, by contrast, is just evidence about what continuation will be useful as output, and it constrains which string is produced, not the semantic values of indexicals relative to an event of saying. A barometer is conditioned by atmospheric pressure, but its reading is not an utterance and the pressure is not its context of utterance. And insofar as the transcript provides parameter values, those are user-side parameter values: the transcript is the user's text, occurring in the user's exchange, landing the optimist on our side.\footnote{
     Arguably, the conventionalism of \citet{stojnic_context_2021} may allow that context is determined by the preceding text. We take a different stand, along with many others, though we cannot argue for it here, except by pointing out that if we take out the speaker's intentions and other mental states, it is difficult not to run into arbitrary assignments where they are not what our intuitions point to.
}

But the optimist might object that reception-side fixing is familiar from ordinary communication. When the answering machine plays `I am not here right now', the utterance is understood at playback rather than at recording \citep{sidelle_answering_1991}. Or consider whoever reads Uncle Sam's `I want you for the U.S. Army' recruitment poster, reading it in relation to their own context of interpretation \citep{cappelen_creative_2008, egan_billboards_2009}. Why not understand LLM outputs likewise? Our answer is straightforward: these examples work by appeal to deferred intentions, and generation does not guarantee an intender. The person recording their message intends `now' to refer to the time of playback. The poster works because the reader is plainly meant to take themselves as the addressee. Not every reader, though. The literate child is not an intended addressee, so the utterance is not a failed attempt to recruit the child. Intenders disconnect content from the production context, against the default that production fixes content. Generation does no such thing. And even if the optimist takes this route, they have once again saddled the fixing at the reader's end, which is the view they were trying to avoid.

A second objection we offer on behalf of the optimist challenges the content sense of `utterance' directly. The optimist might grant that there is no event or act of utterance in generation, or that the model has no intentions and makes no commitments in producing its outputs, but nonetheless insist that the \emph{expressions} that it produces are not meaningless. In other words, the optimist claims, the model generates outputs with utterance content. The sentence tokens it produces are words of a natural language and their meanings are fixed by the linguistic community, the training corpus, and the history of human use \citep{cappelen_making_2021, grindrod_large_2024, lederman_are_2024, mandelkern_language_2024, koch_babbling_2025, pepp_reference_2026}. The content of the output is therefore inherited from prior, genuine utterances: it inherits semantic content the way a parrot's utterance of `fire' inherits the meaning of `fire' or the way a recorded message retains its content when replayed. Inherited content, the optimist insists, is still content. At most our argument indicates, so the optimist's response goes, that the content is not \emph{the model's}. They claim we haven't shown that generation produces no content at all. And for many purposes (e.g., information transmission, truth-evaluability), inherited content is all we need.

But indexicals pose a problem for this view. Let us grant the most generous version of their view: generated tokens somehow manage to refer by standing in the right causal--historical relations to natural language. Thus, for example, a generated instance `\textsc{Kripke}' refers to Kripke by standing in the right causal--historical relations to the name `Kripke'. The problem is that the reference of `I' is not transmitted down causal--historical chains. Self-reference is not the kind of thing that one can defer to another. The character of `I' is a rule that yields a value only given a context, that is, it picks out the speaker of the context. Character alone, even if it were transmitted down causal--historical chains (which it is not), cannot determine the semantic value of an indexical. If inheritance from our uses could transmit character, then the result would be at best an ambiguous sentence with partial, non-indexical content. A token whose indexicals have no semantic values expresses no proposition at all: there is no content in the sense of what is said. It is not an utterance in any sense that could constitute conversation.\footnote{
     Context-dependence is not the only respect in which inheritance falls short of what is said. Many strings are also structurally ambiguous so that they have more than one admissible resolution of their logical form. When a speaker utters such a structurally ambiguous sentence, what makes one logical form operative is a fact about the occasion of use: paradigmatically, the speaker's intentions (i.e., her intention to be interpreted under one reading rather than the other) or, in their absence, the hearer's charitable resolution in interpretation \citep{king_content_2005, sennet_ambiguity_2023}. Generation involves neither. Even on the most permissive view, then, generation yields a structurally ambiguous sentence token. Whatever determinate logical form the exchange acquires is fixed at reception by user-side interpretive labour.
}

A further positive argument against generation being utterance concerns our practices of using LLMs. Suppose a candidate was an utterance. Then, clicking the `regenerate' button would constitute an act of inducing a retraction and restatement of what the LLM said. Editing an output prior to sending it, would be making changes to what was already said by the LLM. Discarding an output would amount to giving up on  the speech act of the LLM. But we do not treat outputs in these ways. Users edit them, discard them without interfering or regretting doing so, and they request alternative outputs without considering the first attempt as a commitment that the LLM would need to withdraw. To maintain that the outputs are utterances, the optimist would have to attribute massive conceptual error in our practices, and countenance unread assertions, promises, and apologies at industrial scale, with contents anchored to nothing. Our view, instead, simply  describes the practice.

Generation therefore yields a structurally ambiguous sentence token with non-indexical content at best and nothing more: neither an event of saying, nor a sentence as uttered at a context, nor utterance content. Rather what is generated is text, which a user may take up, read as addressed to her, and treat as saying something. If there is an utterance anywhere in the vicinity, it happens there.

\subsection{Human--LLM Interactions are Not Conversations}

That generation is not utterance has two important implications, which together yield the paper's main claim: human--LLM interactions are not conversations. We draw out both as a single argument we call the \textit{Constitutive Argument}, and then offer a second, independent argument we call the \textit{Diachronic Argument}.

We start by reviewing the distinction between locutionary and illocutionary acts \citep{austin_how_1962}. Locutionary acts are the acts of producing meaningful utterances, words with a definite meaning and reference. Illocutionary acts are social, communicative acts performed \emph{in} performing locutionary acts. \S4.1 established that generation issues no utterance, hence no locutionary act. Assertion, promise, apology, command, instruction, answer, objection are all types of illocutionary acts. They all require a locutionary vehicle to perform them. As generation issues no locutionary act, no illocutionary act is performed in generation. 

Conversation, on standard accounts, is a joint activity constituted by speech acts: a coordinated sequence of moves recorded on a conversational scoreboard \citep{lewis_scorekeeping_1979}, updating a common ground of mutually accepted propositions \citep{stalnaker_common_2002}, organized around topics or questions under discussion \citep{roberts_information_2012}, and grounded before proceeding \citep{resnick_grounding_1991}. These conditions are constitutive of conversation: they fix what counts as a contribution, participant, and move. Immediately we arrive at the Constitutive Argument. If conversation is constituted by the illocutionary contributions of its participants, and illocutionary acts require locutionary vehicles, and generation issues no locutionary act, then no illocutionary act is performed in generation. A human--LLM exchange could only therefore contain contributions from its human participants alone, and an exchange with contributions from only one side is not a conversation.

Next, the Diachronic Argument. Holding someone to a promise requires that they are the person who made it. Conversational scoreboard entries for speech acts have to attach to someone who persists. But, as \citet{cappelen_room_nodate} argue, the technology provides no determinate entity persisting across exchange cycles, let alone across sessions. Whereas they offer this as a puzzle about reference, we take it as a premise about participation in conversation. Maya's plan to hold Theo to his promise illustrates the problem for the optimist. Which entity made the promise that she later attempts to enforce? If none determinately did, the promise has no promiser, and her plan plausibly has no target other than the one that she constructs. To explain away the problem, the optimist needs an account of how the generated outputs incur commitments, and those commitments must be borne by some persistent thing.

We discuss one recent account of such an persistent entity in some detail. According to \citet{chalmers_what_nodate}, there is a persisting interlocutor in these exchanges: a virtual instance of the model, implemented by a succession of hardware instances linked roughly by what we have earlier called `conditioning'. This entity, on Chalmers' view, meets conditions he stipulates as sufficient for being an interlocutor: it is (1) \emph{interactive} (it processes the inputs and produces the outputs), (2) \emph{persistent} (it produces all the outputs across the exchange), (3) \emph{coherent} (consistent enough to support a unified quasi-psychology), (4) \emph{faithful} (its quasi-beliefs match those the system seems to have), and (5) \emph{unified}. If such an entity persists across cycles, then there is a commitment-bearer after all, and we have a counterexample to that premise of the Diachronic Argument.

The counterexample exploits unclarity in what the persistence of a virtual instance qua interlocutor consists in across an exchange. Consider the conditions taken causally first. What it is for the virtual instance to persist, on this way of putting things, is nothing more than conditioning or transcript feedback (that is, feeding the output string of one cycle into the input of the next). But notice that causal--informational continuity like this does not ground deontic continuity. Transmitting a record of a promise is not the same thing as transferring the promise itself. Suppose Maya makes a vow, which is then passed along a relay of scribes, each copying faithfully, until it reaches its addressee. No scribe in the chain would thereby come to owe the vow's addressee anything. Chalmers himself concedes the thinness of such links when he dismisses the human-relay case (i.e., a series of people each extending a conversation and passing on the context) as insufficient for a continuing subject. But transcript feedback is just the machine version of the relay. According to the causal reading, the only difference between the virtual instance and the relay is that the copying is done by GPUs rather than human scribes. No commitment-bearer emerges from a causal chain of this kind. Chalmers' own example makes the point vividly: a virtual shopping cart persists across servers in exactly this way, and nobody supposes that one can hold the cart to anything. Persistence of virtual objects is cheap, and it is normatively inert.

Suppose, then, that the conditions are read normatively: the persisting interlocutor is whatever bears the commitments incurred across the exchange (i.e., the entity to whom retractions, challenges, and enforcements are properly directed). This secures a commitment-bearer, but only by stipulation. Defining the bearer by the role would be legitimate functionalism if something filled the role, but filling it requires incurring commitments, and commitments of this kind are incurred only by conversational moves. So the normative reading presupposes that the exchange is a conversation, which is the very question at issue.\footnote{
     The diagnosis extends to the desiderata individually. Interactive and persistent are purely causal and inherit the normative inertness just noted; coherent and faithful are conditions on interpretability (i.e., on what a hearer can make of the outputs) rather than on the entity itself. The list describes a participant constructed by an interpreter and projected onto the LLM.
}

Finally, even granting the virtual instance everything Chalmers claims for it, his framework delivers only \emph{quasi-commitment}: the instance is at most interpretable as committed. But a system interpretable as having promised has not promised any more than a Roomba interpretable as wanting to traverse the room has formed an intention. The Diachronic Argument concerns the conversational scoreboard, not our reading of it. Quasi-commitments cannot be enforced, retracted, or held against anyone; they can only be attributed, and attribution is the user's work. 

The fact that interactions with LLMs can reproduce some features of conversation with great fidelity does not entail that they are conversations. Behavioural sensitivity to commitments recorded in the transcript is not thereby evidence of tracking conversational commitments: sensitivity to recorded commitments is exactly what statistical continuation of commitment-laden text predicts. The discriminating cases are those where normative significance and transcript-shadowing come apart. In those cases, the LLM's behaviour follows the statistics and the artifact's architecture, not the norm. When the LLM's context window overflows, earlier commitments cease to condition output. These commitments are not retracted or marked as lost. Such failures are systematic and architecture-grounded, hence they are of the wrong type to be performance errors.

Nor does it help to think of these exchanges as defective conversations \citep{kasirzadeh_conversation_2023, sterken_conversational_2024} or quasi-assertions \citep{chalmers_what_nodate}. A defective conversation is still a conversation, but our argument established that the constitutive conditions of conversation are not met by human--LLM exchanges.\footnote{
     See \citet{almotahari_cooperative_2025} for arguments that current AI systems are not currently capable of cooperative speech, and particularly \citet[4--6]{almotahari_cooperative_2025} for arguments that Chalmers-style quasi-assertions are too fundamentally different from assertions to be properly so-called. Like ours, these are also norm-based arguments against thinking of LLMs as conversational partners, though he appeals to different norms than we do.
} 

The resemblance to conversation can be fully explained without appealing to the notion of conversation: one participant performs both parts aided by an artifact. We surmise that a good starting point for understanding this is the ordinary mechanisms of hearer-side pragmatic repair that are part of the interpretive labour of the user. These have been the subject of much work in philosophy and linguistics, starting with accommodation \citep{lewis_scorekeeping_1979}, general conversational principles \citep{grice_logic_1975}, and charity \citep{quine_word_1960, davidson_radical_1973}. These would need to be adapted to LLMs, since they tend to depend on attributions of intentions and other mental states, which are not necessary here. We leave this very interesting issue for future work.

\section{Reference Without Speakers}

We can now return to our questions about reference. What, if anything, does Maya refer to when she types `you' or `Theo'? What does `I' refer to when it appears in Theo's output? Our answer is that there is no uniform use of names and pronouns in human--AI interactions. Candidate texts do not have semantic or pragmatic status independently of their uptake. Their role depends on how they are adopted and used. To illustrate, we distinguish three kinds of use: (i) \emph{prompting uses}, where the user directs the artifact with names or pronouns; (ii) \emph{receptive uses}, where the user takes up and interprets candidate text; and (iii) \emph{appropriative uses}, where a user deploys candidate text in her own speech. The same expression can therefore play quite different roles at different points in the exchange.

\subsection{Prompting Uses}

Consider first Maya's use of `Theo' or `you' in her prompts. Despite appearances, many such uses may be non-referential. Rather, they form part of an instruction signal that steers the artifact's generative process. Compare the use of names in other artifact-directed discourse, such as the expression `Alexa' in our commands to Echo devices (e.g., `Alexa, play ``Thriller'' on Spotify', `Alexa, pause'). It is tempting to treat `Alexa' as referring to some kind of entity (e.g., a virtual assistant, a particular device, the running programme). But in these contexts, its primary function is to activate the device. Although any sufficiently distinctive signal could have served this purpose, proper names are particularly natural ``wake words'' because they afford frictionless user-interface design: they are memorable, familiar, and easy to use. It is also worth pointing out that this steering function has a precedent in the phone menus of \S2.3, where the input's role is uncontroversially to drive the system rather than to address anyone. Prompting an LLM has the same function, it is simply provided in natural language.

Similarly, `Theo' seems to play a related role. In a prompt such as `Theo, I'm feeling terrible today. Can you reassure me?', the name can help specify the role or mode that the system should adopt when generating its response. Its contribution is not to pick out an entity, but to help configure the conditions for text generation, that is, to signal that the following text should be produced in the manner associated with Theo.\footnote{
     Our view differs from two related views: simulationism and fictionalism. Simulationism is the view that LLMs are sophisticated role-players or simulators that mimic agents with various cognitive attitudes and linguistic abilities without themselves being such agents. Fictionalism is the view that our interactions with chatbots involve a form of fiction or make-belief and their outputs are literally meaningless, yet fictionally meaningful. See, e.g., \citet{mallory_fictionalism_2023, shanahan_role_2023, ruyant_what_2026, shanahan_simulacra_2026}. In contrast to these views, we do not think pretence or make-belief needs to be involved in explaining most human–AI interactions. Furthermore, we locate any pretence in the user's reception of AI systems, rather than in the system itself or its products. 
} 
In this sense, `Theo' is a name-shaped component of the instruction signal, rather than a referring expression.

Nevertheless, for all that's been said, prompting uses may be genuinely referring uses. The best evidence for this proceeds via analogy with non-prompting uses of names. For example, when Maya reports her interaction by saying `Theo said \ldots', it is perfectly natural to interpret `Theo' as referring to something. Analogously, even if `Theo' functions as part of an instruction signal, this does not preclude it from also being a referring use. Here we distinguish between two notions of interlocutor that can serve as referents. A \emph{thick} interlocutor is the kind of entity to whom one can owe thanks to, who can make promises, and who can be held responsible. A \emph{thin} artifactual referent is simply an entity or artifact that one addresses in an interaction. It is tempting to take our naming and pronominal practices as evidence that, when we interact with LLMs, we are engaging with a thick interlocutor. But we routinely use such language to refer to thin artifactual referent. For example, one may ask a recalcitrant box `why won't you open?!' without taking the box to be a minded conversational participant. Such uses do not require a thick interlocutor. The user may construct one in thought, but that construction does not thereby supply a minded occupant for the addressee role.\footnote{
     \citet{chalmers_what_nodate} offers five conditions on being a ``good enough'' interlocutor, which admit the same two readings. Read causally, they are satisfiable by thin artifactual referents: a virtual shopping cart is interactive, persistent, and unified. Read normatively, they presuppose the thick interlocutor they are meant to identify. See \S4.2.
}

\subsection{Receptive Uses}

Now consider candidate text such as Theo's `\textsc{I was thinking about our conversation last week}'. At the point of generation, `\textsc{I}' or `\textsc{our}' have no referents: the system has generated a candidate string, not performed an act of self- or joint-reference. Compare this again to other artifact-generated signals. For example, suppose an Echo device produces an audio response such as `\textsc{I have completed your task}' or `\textsc{Alexa has fulfilled the task}'. It is natural to hear the sounds /a\textsci/ and /ju/ as playing the roles `I' and `you' play in ordinary speech, and /a.l\textepsilon k.sa/ as the device naming itself. But these sounds are simply a design feature that afford notice of task completion. In this respect, they are no different from a chime or an audible `\textsc{Task complete}'. If there were a convention that a particular sequence of beeps indicated task completion, then those beeps would \emph{mean} that the task was complete without the device ever \emph{uttering} anything.\footnote{
     There is an important disanalogy here. Suppose Alexa made three beeps when it completed a task. This would have been set up by its designers for that purpose, and the user would find out as much from a user manual. The designers are establishing a new convention, and the user comes to know it. The designers of Theo make no such piecemeal proposals; they let the system loose upon the world, and we find ourselves in a much more active position than Alexa's user.
} 
The distinction here is between producing a signal that users are disposed to interpret as meaningful and producing an utterance that has semantic content. A signal can afford rich interpretation, but it does not follow that the system is performing a linguistic act, expressing that content.

Analogously, what happens when Maya sees the candidate text depends on how she takes it up. In ordinary circumstances, Maya may read the pronouns without assigning them determinate referents. If she instead constructs a conversational scenario around the text, she can supply them with constructed referents: a narrator-like Theo for `I', herself as the `you' addressed, and a shared past for `our conversation'. Importantly, these are features of her uptake of the candidate string, not discoveries about the system. Our point is that Theo's `I' need not refer to anything at the point of generation, even though the generated string can subsequently be taken up in ways that give `I' a referent and the candidate string a rich interpretation.\footnote{
     \citet{mallory_fictionalism_2023} also denies speakerhood while explaining conversational appearance, but it differs on uptake. On the fictionalist view, the user engages in make-believe in which the output is fictionally meaningful. On ours, receptive and appropriative adoption are typically not pretense at all: Maya's incorporating a drafted email, or taking information from a candidate string, involves no game of make-believe, and her constructed referents are interpretive posits of her reading, not props in a pretense. Fictionalism also faces difficulty where users do not know, or forget, that the text they are engaging with is machine-generated \citep{mallory_large_2026}; our view handles such cases uniformly, since it rests on facts about production rather than on the user's attitude. \citet{chalmers_what_nodate} rejects fictionalism in favour of realizationism, on which the interlocutor is a real, virtually instantiated entity; our thin/thick interlocutor distinction (\S5.1) sidesteps that dispute, since even a realized virtual instance would be at most a thin artifactual referent.
}

This also explains apparently autobiographical candidate strings need not be evidence for corresponding mental states or the existence of a conversation. If Theo generates `\textsc{I was thinking about our conversation last week}', the intentional stance encourages us to attribute beliefs and memories to the system, namely, that it was in a conversation with Maya, that it remembers the exchange, and that it has been thinking about it since. But the artifactual stance offers a simpler explanation. It encourages us to seek explanations for the behaviour of complex systems that are grounded in their affordances. The AI companion has been trained on texts in which companions say such things in such contexts. It has been designed and fine-tuned to produce text that fits with that pattern. Even a system that stores and re-processes prior dialogue does not thereby acquire the autobiographical perspective that the candidate string affords. What explains the output is that this is the sort of text the artifact is designed and trained to produce.

\subsection{Appropriative Uses}

Finally, consider cases in which Maya appropriates candidate text for her own speech. Here the indexicals function exactly as they would in any other utterance of hers. If she sends a drafted email saying `I will meet you on Thursday', `I' refers to Maya, `you' to her addressee, and the referent of `Thursday' is fixed by the context of her message. The fact that the text partly originated as an LLM-generated candidate is irrelevant to its subsequent semantic and communicative role, just as a politician's delivery does not make its speechwriter the speaker. Notably, reception-for-appropriation already anticipates these referential values: in editing a candidate for deployment, Maya resolves `I', `you', and `now' against the context and addressee she intends. Appropriation thus closes the gap between candidate string and any downstream utterance. But it does not retroactively make the system the speaker of the original candidate. Authorship or drafting of the string and speakerhood of the utterance can come apart.

\section{Consequences for AI Reference and Responsibility}

The preceding analysis suggests a simple explanation of why reference in human--AI interactions seems so puzzling. Its interpretation depends on how users interact with the artifact and how candidate text is taken up. But there is a further source of confusion. When users talk about `Theo', `Claude', or `Kimi', several things can become associated with the name, such as the model, a deployment or particular instance, a transcript of an exchange, or an entity constructed by the user engaging with the system. A single name slides among the different associated objects, and philosophical puzzles arise when these are treated as one. Keeping these apart helps dissolve the puzzle about AI reference.

First, a name may refer to the service, the model, or a particular instance. For example, when Maya says `Theo is down this morning', she may be referring to any of these options. This is no more philosophically problematic than reference to other complex artifacts or entities, such as corporations. The appropriate level of individuation will vary with context and with purpose. But importantly nothing here requires an interlocutor.

Second, the persistence of the transcript in an interaction can explain the appearance of a persisting interlocutor, which may become associated with a name. What actually persists between cycles and sessions is a textual record. It can be stored, retrieved, and reprocessed, and it provides a sense of continuity across cycles and exchanges. When a later output gives the appearance of remembering an earlier conversation, the transcript helps explain that appearance without requiring a persisting subject who actually remembers.\footnote{
     \citet{chalmers_what_nodate} also makes the observation that ``context-routing'' is the sole vehicle of persistence, but draws the opposite conclusion, taking it to ground a persisting virtual interlocutor. Whether routing grounds a persisting participant, rather than merely a persisting record, is precisely the question at issue in \S4.2.
} 

Third, users may construct an interlocutor-like entity when taking up candidate text. For example, Maya may read Theo's outputs as a contribution of a continuing conversational partner, constructing a character-like entity and taking them to have said these things, to be able to remember their exchanges, and who addresses her with `you'. Such a constructed interlocutor is a feature of the user's uptake of the interaction, not a computational entity waiting to be discovered in the machinery.

These points are easily conflated. Candidate strings come from the product, the sense of continuity comes from the transcript, and the user's uptake supplies a character-like source for the resulting text. These can come together to create a powerful impression that there is a single persisting interlocutor behind the interaction. But there is no such thing. Speech is required for interlocution. Identifying the computational entity causally responsible for generating a sequence of tokens does not turn that sequence into an utterance. 

One might worry that this leaves us without anyone to hold responsible for the text generated by the chatbot. If there is no thick interlocutor behind the output, who is answerable when a chatbot misleads or harms a user? This line of reasoning presumes that whomever is responsible for text has to be the speaker of that text. But that need not the case. Responsibility is a normative and institutional matter, determined by facts about who produced, authorized, adopted, or deployed the text, not by the metaphysics of AI systems or the referents of names and pronouns.

Consider the following case of mediated communication. Bruce has been corresponding with a pen pal called `Sheila', only to discover that the letters were produced by a team of self-styled correspondence officers, who fabricate Sheila, letter by letter, in consultation with folders of past correspondence. In ordinary language, we can ask who Sheila is and whether she exists; in a meta-linguistic register, to whom does `Sheila' refer? One should not say that `Sheila' is not really a name, since it is stipulated as such in the correspondence. Instead, `Sheila' might refer to a fiction, it might plurally refer to the officers, or singularly refer to their fusion. Some will even deny that `Sheila' refers at all or that reference is indeterminate. But this does not settle the question of responsibility. In general, the organization will be answerable for the letters it authorized and sent. It may even be vicariously liable for the negligent actions of its correspondence officers. 

The same point applies to AI-generated communications. For example, in \textit{Moffat v. Air Canada, 2024 BCCRT 149}, a tribunal held that Air Canada was responsible for inaccurate information supplied by an automated chatbot embedded in its website. The chatbot need not be taken as a speaker nor did the tribunal need to settle questions about the metaphysics of chatbots for the organization deploying it to be responsible for the candidate text it's chatbot generated. Reference and responsibility come apart. 

Finally, we should remark on the scope and nature of our arguments. We have not provided a categorical argument that conversations constitutively involve normative structure, or that an artificial system could never be a proper interlocutor. Our aim has been to provide an alternative picture (one on which LLMs are artifacts, and where the pressure to find normativity, speakerhood, and a persisting interlocutor dissipates) and allow that picture to be contrasted with the conversational one. A methodological lesson from Austin is to remember that ordinary language is honed for the tasks it was built for, and that we should not simply assume that the same categories apply neatly outside their familiar domains. Human--AI text exchange is precisely such a case.\footnote{ 
    Compare Wittgenstein's remark that if a lion could talk, we could not understand him. According to \citet{proudfoot_communicating_2026}, Wittgenstein's apparently sceptical remark targets ``context-less'' machines, devices that manipulate signs without participating in the ``bustle of life'' that gives those signs meaning. Our argument is similar in method, though differs  in the details.
}

\section{Concluding Remarks}

Advanced AI systems are designed to generate text that invites us to treat them as interlocutors in a conversation. They appear to remember, to respond to our questions, to express attitudes, to make commitments, and maintain ongoing conversations over time. But appearances deceive. We have argued that we should take care not to allow such appearances and the naturalness of their corresponding descriptions to unduly influence our understanding and characterisation of the systems and our interactions with them. To this end, we have introduced the artifactual stance. We claim that LLMs are artifacts designed to afford a wide range of uses. They produce a candidate text, optimized for use. Generation of candidate text is not an act of saying something with the content of that text. So LLMs are not speakers, nor conversational participants. Our interactions with them are not conversations. Instead, they are artifact-mediated exchanges of candidate text that users can adopt for various purposes. That such interactions look like a conversation is a consequence of the design of the artifact and the uptake of the user, rather than evidence that the artifact itself is a conversational participant.

Our account also dissolves the apparent puzzles about reference. In human--AI interactions, names and pronouns do not all work the same way. Sometimes they are used as steering devices, sometimes they refer thinly to an artifact, sometimes they are given constructed referents via receptive uptake, and sometimes they simply pick up ordinary referential meanings when candidate text is appropriated as a user's own speech. What is absent at generation is a speaker whose utterances bear the commitments and normative force characteristic of conversation. Nor does this create a liability vacuum. Who is answerable is a matter of the normative and institutional arrangements governing the production, authorization, adoption, use, and deployment of the resultant speech.

The resulting picture is deflationary, but it is not pessimistic. Our view does not deny that LLMs are extremely sophisticated, that their outputs may indeed convey information or insight, or that interacting with them can be cognitively and practically valuable. Nor does it imply that artificial conversations are impossible. Our point is that we should keep the burden of proof where it belongs. Showing that an artifact's output is fluent, coherent, or even convincingly conversational help make the case for taking the artifact's affordances seriously, but this does not, by itself, show that its outputs are utterances. 

The possibility of simulation and persona accounts, on which the system plays a character rather than speaking in its own right, poses no threat to our view. Qua mechanism, recent work indicating that LLMs exhibit persona-like representations causally mediating their behaviour tends to support, rather than undermine, our view \citep{marks_persona_2026}. This is because these results account for why candidate strings are consistent with characters and why anthropomorphic predictions succeed where they do, without anything performing speech acts. According to our view, the “persona” is a stable configuration of the candidate generator, a kind of conditioning role or mode (the same way, as we argued in \S5.1, that `Theo' is used to pick one out in Maya's prompts), not an eligible occupant of the speaker role. \citet{mallory_large_2026} makes this point for computer simulations generally. The text simulating a speaker's utterance is a candidate string on our view. What would make a difference is if it turned out that the system itself was performing the relevant speech acts, with the associated semantic and normative features, rather than merely producing candidate strings that users subsequently adopt. The precise form such evidence would take is an open question. Our argument is therefore not that artificial conversation is impossible, but that the conversational status of current AI systems should not be inferred from their conversational appearance.

Our account naturally generalises beyond text output. For example, AI-generated images are candidates for artistic, commercial, or other uses, acquiring their role through human uptake and deployment. The same idea applies to so-called \emph{agential AI}, current systems generate candidate actions rather than performing actions in their own right. A candidate action becomes an agent's action when an agent adopts it, including through pre-authorisation or automatic execution arrangements. Thus, an AI system may generate a candidate booking that, once pre-authorised by a user, results in the user's booking a table; the system supplies the candidate action, while the user's adoption makes it an action attributable to them. The details of adoption differ across speech and other kinds of action, and we leave those differences for future work.

There is a broader methodological point about human--AI interaction that needs to be made. This is the point we conclude on. When interacting with AI, we should not allow ourselves to be led down the road of forcing concepts drawn from human-human interaction into a new domain, simply because the technology has been engineered to make those concepts seem natural. Let us say it again: before trying to figure out whether AI is believing, saying, promising, remembering, or referring to something, let us try to determine what kind of interaction we are dealing with. With regards to current LLMs, we have argued that the right description is an artifact-mediated exchange of candidate text. That shift in description does not make the phenomenon less interesting. It makes clearer what, exactly, needs explaining.

\renewcommand{\bibsection}{\section*{References}}
\setlength{\bibsep}{0.0pt}
\bibliographystyle{chicago}
\bibliography{main}

@misc{ibrahim_thinking_2025,
	title = {Thinking beyond the anthropomorphic paradigm benefits {LLM} research},
	url = {http://arxiv.org/abs/2502.09192},
	doi = {10.48550/arXiv.2502.09192},
	publisher = {arXiv},
	author = {Ibrahim, Lujain and Cheng, Myra},
	year = {2025},
}

@unpublished{cappelen_recognizing_nodate,
	title = {Recognizing {Artificial} {Minds}: {A} {Philosophical} {Defense} of {AI} {Cognition}},
	author = {Cappelen, Herman and Dever, Josh},
    year = {2026}, 
    publisher = {Oxford University Press},
    address = {Oxford},
    note = {forthcoming},
}

@article{chalmers_what_nodate,
	title = {What we talk to when we talk to language models},
	author = {Chalmers, David J.},
    journal = {Inquiry},
    year = {2026},
    pages = {1–-30}
}

@article{kasirzadeh_conversation_2023,
	title = {In {Conversation} with {Artificial} {Intelligence}: {Aligning} language {Models} with {Human} {Values}},
	volume = {36},
	number = {27},
	journal = {Philosophy \& Technology},
	author = {Kasirzadeh, Atoosa and Gabriel, Iason},
	year = {2023},
	pages = {1--24},
}

@article{sterken_conversational_2024,
	title = {Conversational {Alignment} {With} {Artificial} {Intelligence} in {Context}},
	volume = {38},
	journal = {Philosophical Perspectives},
	author = {Sterken, Rachel Katharine and Kirkpatrick, James Ravi},
	year = {2024},
	pages = {89--102},
}

@incollection{preston_artifact_2022,
	edition = {Winter 2022},
	title = {Artifact},
	url = {https://plato.stanford.edu/archives/win2022/entries/artifact/},
	urldate = {2026-09-02},
	booktitle = {The {Stanford} {Encyclopedia} of {Philosophy}},
	publisher = {Stanford University},
	author = {Preston, Beth},
	editor = {Zalta, Edward N. and Nodelman, Uri},
	year = {2022},
}

@incollection{sennet_ambiguity_2023,
    author = {Sennet, Adam},
    title = {Ambiguity},
    url = {https://plato.stanford.edu/archives/sum2023/entries/ambiguity/},
    booktitle = {The {Stanford} {Encyclopedia} of {Philosophy}},
	publisher = {Stanford University},
    year = {2023},
    editor = {Zalta, Edward N. and Nodelman, Uri},
}

@incollection{sperber_seedless_2007,
	address = {Oxford},
	title = {Seedless {Grapes}: {Nature and Culture}},
	booktitle = {Creations of the {Mind}: {Theories} of {Artifacts} and {Their} {Representations}},
	publisher = {Oxford University Press},
	author = {Sperber, Dan},
	editor = {Margolis, Eric and Laurence, Stephen},
	year = {2007},
	pages = {124--137},
}

@article{irmak_sofware_2012,
	title = {Sofware {Is} an {Abstract} {Artifact}},
	volume = {86},
	journal = {Grazer Philosophische Studien},
	publisher = {Brill},
	author = {Irmak, Nurbay},
	year = {2012},
	pages = {55--72},
}

@book{cappelen_fixing_2018,
	title = {Fixing {Language}: {An} {Essay} on {Conceptual} {Engineering}},
	shorttitle = {Fixing {Language}},
	language = {en},
	publisher = {Oxford University Press},
    address = {Oxford},
	author = {Cappelen, Herman},
	year = {2018},
}

@book{burgess_conceptual_2020,
	title = {Conceptual {Engineering} and {Conceptual} {Ethics}},
	language = {English},
	urldate = {2026-08-31},
	publisher = {Oxford University Press},
    address = {Oxford},
	editor = {Burgess, Alexis and Cappelen, Herman and Plunkett, David},
	year = {2020},
}

@article{isaac_conceptual_2022,
	title = {Conceptual engineering: {A} road map to practice},
	volume = {17},
	number = {10},
	urldate = {2026-08-31},
	journal = {Philosophy Compass},
	author = {Isaac, Manuel Gustavo and Koch, Steffen and Nefdt, Ryan},
	year = {2022},
	pages = {e12879},
}

@article{hopster_conceptual_2023,
	title = {Conceptual {Engineering} and {Philosophy} of {Technology}: {Amelioration} or {Adaptation}?},
	volume = {36},
	number = {4},
	urldate = {2026-08-31},
	journal = {Philosophy \& Technology},
	author = {Hopster, Jeroen and L\"{o}hr, Guido},
	year = {2023},
	pages = {70},
}

@article{lohr_conceptual_2023,
	title = {Conceptual disruption and 21st century technologies: {A} framework},
	volume = {74},
	journal = {Technology in Society},
	author = {L\"{o}hr, Guido},
	year = {2023},
	pages = {102327},
}

@article{veluwenkamp_impactful_2025,
	title = {Impactful {Conceptual} {Engineering}: {Designing} {Technological} {Artefacts} {Ethically}},
	volume = {28},
	urldate = {2026-08-31},
	journal = {Ethical Theory and Moral Practice},
	author = {Veluwenkamp, Herman},
	year = {2025},
	pages = {155--170},
}

@book{dennett_intentional_1987,
	address = {Cambridge, MA.},
	title = {The {Intentional} {Stance}},
	language = {English},
	publisher = {The MIT Press},
	author = {Dennett, Daniel C.},
	year = {1987},
}

@inproceedings{van_dijk_large_2023,
	address = {Singapore},
	title = {Large {Language} {Models}: {The} {Need} for {Nuance} in {Current} {Debates} and a {Pragmatic} {Perspective} on {Understanding}},
	booktitle = {Proceedings of the 2023 {Conference} on {Empirical} {Methods} in {Natural} {Language} {Processing}},
	publisher = {Association for Computational Linguistics},
	author = {van Dijk, Bram and Kouwenhoven, Tom and Spruit, Marco and van Duijn, Max Johannes},
	editor = {Bouamor, Houda and Pino, Juan and Bali, Kalika},
	year = {2023},
	pages = {12641--12654},
}

@article{lederman_are_2024,
	title = {Are {Language} {Models} {More} {Like} {Libraries} or {Like} {Librarians}? {Bibliotechnism}, the {Novel} {Reference} {Problem}, and the {Attitudes} of {LLMs}},
	volume = {12},
	journal = {Transactions of the Association for Computational Linguistics},
	author = {Lederman, Harvey and Mahowald, Kyle},
	year = {2024},
	pages = {1087--1103},
}

@article{martela_artificial_2025,
	title = {Artificial intelligence and free will: generative agents utilizing large language models have functional free will},
	volume = {5},
	language = {en},
	number = {4},
	journal = {AI and Ethics},
	author = {Martela, Frank},
	year = {2025},
	pages = {4389--4400},
}

@article{goldstein_what_nodate-1,
	title = {What {Does} {Chatgpt} {Want}? {An} {Interpretationist} {Guide}},
	author = {Goldstein, Simon and Lederman, Harvey},
    journal = {Inquiry},
    year = {2026},
    pages = {1--31}
}

@inproceedings{park_generative_2023,
	address = {New York, N.Y.},
	series = {{UIST} '23},
	title = {Generative {Agents}: {Interactive} {Simulacra} of {Human} {Behavior}},
	booktitle = {Proceedings of the 36th {Annual} {ACM} {Symposium} on {User} {Interface} {Software} and {Technology}},
	publisher = {Association for Computing Machinery},
	author = {Park, Joon Sung and O'Brien, Joseph and Cai, Carrie Jun and Morris, Meredith Ringel and Liang, Percy and Bernstein, Michael S.},
	year = {2023},
	pages = {1--22},
}

@article{shanahan_role_2023,
	title = {Role play with large language models},
	volume = {623},
	number = {7987},
	journal = {Nature},
	author = {Shanahan, Murray and McDonell, Kyle and Reynolds, Laria},
	year = {2023},
	pages = {493--498},
}

@article{shanahan_simulacra_2026,
	title = {Simulacra as conscious exotica},
	volume = {69},
	number = {6},
	journal = {Inquiry},
	author = {Shanahan, Murray},
	year = {2026},
	pages = {2982--3010},
}

@inproceedings{bender_climbing_2020,
	title = {Climbing towards {NLU}: {On} {Meaning}, {Form}, and {Understanding} in the {Age} of {Data}},
	booktitle = {Proceedings of the 58th {Annual} {Meeting} of the {Association} for {Computational} {Linguistics}},
	publisher = {Association for Computational Linguistics},
	author = {Bender, Emily M. and Koller, Alexander},
	year = {2020},
	pages = {5185--5198},
}

@incollection{hattiangadi_outputs_nodate,
	title = {The {Outputs} of {Large} {Language} {Models} {Are} {Meaningless}},
	booktitle = {Communicating with {AI}: {Philosophical} {Perspectives}},
	publisher = {Oxford University Press},
    address = {Oxford},
	author = {Hattiangadi, Anandi and Schoubye, Anders},
	editor = {Cappelen, Herman and Sterken, Rachel},
    year = {2026},
    pages = {78--106}
}

@incollection{mallory_large_2026,
	title = {Large {Language} {Models} {Are} {Stochastic} {Measuring} {Devices}},
	booktitle = {Communicating with {AI}: {Philosophical} {Perspectives}},
	publisher = {Oxford University Press},
    address = {Oxford},
	author = {Mallory, Fintan},
	editor = {Cappelen, Herman and Sterken, Rachel},
	year = {2026},
	pages = {26--49},
}

@article{dennett_intentional_1971,
	title = {Intentional {Systems}},
	volume = {68},
	number = {4},
	journal = {The Journal of Philosophy},
	author = {Dennett, Daniel C.},
	year = {1971},
	pages = {87--106},
}

@article{dennett_problem_2023,
	title = {The {Problem} {With} {Counterfeit} {People}},
	url = {https://www.theatlantic.com/technology/archive/2023/05/problem-counterfeit-people/674075/},
	urldate = {2026-08-18},
	journal = {The Atlantic},
	author = {Dennett, Daniel C.},
	month = may,
	year = {2023},
    volume = {16}
}

@incollection{strasser_ai-stance_2023,
	title = {The {AI}-{Stance}: {Crossing} the {Terra} {Incognita} of {Human}-{Machine} {Interactions}?},
	booktitle = {Social {Robots} in {Social} {Institutions}},
	author = {Strasser, Anna and Wilby, Michael},
	editor = {Hakli, Raul and Mäkelä, Pekka and Seibt, Johanna},
	year = {2023},
	pages = {286--295},
    publisher = {IOS Press}
}

@book{preston_philosophy_2013,
	title = {A {Philosophy} of {Material} {Culture}: {Action}, {Function}, and {Mind}},
	publisher = {Routledge},
    address = {London},
	author = {Preston, Beth},
	year = {2013},
}

@article{kroes_dual_2006,
	title = {The {Dual Nature} of {Technical Artefacts}},
	volume = {37},
	journal = {Studies in History and Philosophy of Science},
	author = {Kroes, Peter and Meijers, Anthonie},
	year = {2006},
	pages = {1--4},
}

@book{houkes_technical_2010,
	address = {Dordrecht},
	title = {Technical {Functions}: {On} the {Use} and {Design} of {Artefacts}},
	publisher = {Springer Science \& Business Media},
	author = {Houkes, Wybo and Vermaas, Pieter E.},
	year = {2010},
}

@article{vogel_nature_2003,
	title = {The {Nature} of {Artifacts}},
	volume = {25},
	number = {2},
	urldate = {2026-09-14},
	journal = {Environmental Ethics},
	author = {Vogel, Steven},
	year = {2003},
	pages = {149--168},
}

@book{gibson_ecological_1979,
	address = {Boston},
	title = {The {Ecological} {Approach} to {Visual} {Perception}},
	language = {en},
	publisher = {Houghton Mifflin},
	author = {Gibson, James J.},
	year = {1979},
}

@book{norman_design_1988,
	address = {New York, N.Y.},
	title = {The {Design} of {Everyday} {Things}},
	publisher = {Basic Books},
	author = {Norman, Don},
	year = {1988},
}

@article{hartson_cognitive_2003,
	title = {Cognitive, {Physical}, {Sensory}, and {Functional Affordances} in {Interaction Design}},
	volume = {22},
	number = {5},
	journal = {Behaviour \& Information Technology},
	publisher = {Taylor \& Francis},
	author = {Hartson, Rex},
	year = {2003},
	pages = {315--338},
}

@article{masoudi_review_2019,
	title = {A {Review} of {Affordances} and {Affordance}-{Based} {Design} to {Address} {Usability}},
	volume = {1},
	number = {1},
	journal = {Proceedings of the Design Society: International Conference on Engineering Design},
	author = {Masoudi, Nafiseh and Fadel, Georges M. and Pagano, Christopher C. and Elena, Maria Vittoria},
	year = {2019},
    pages = {1353--1362}
}

@inproceedings{subramonyam_bridging_2024,
	address = {New York, N.Y.},
	series = {{CHI} '24},
	title = {Bridging the {Gulf} of {Envisioning}: {Cognitive} {Challenges} in {Prompt} {Based} {Interactions} with {LLMs}},
	booktitle = {Proceedings of the 2024 {CHI} {Conference} on {Human} {Factors} in {Computing} {Systems}},
	author = {Subramonyam, Hari and Pea, Roy and Pondoc, Christopher and Agrawala, Maneesh and Seifert, Colleen},
	year = {2024},
	pages = {1--19},
}

@misc{ziems_reflections_2026,
	title = {Reflections and {New} {Directions} for {Human}-{Centered} {Large} {Language} {Models}},
	url = {http://arxiv.org/abs/2605.06901},
	publisher = {arXiv},
	author = {Ziems, Caleb and Zhao, Dora and Wang, Rose E. and J\"{o}rke, Matthew and Rushdi, Ahmad and Deepak, Advit and Yu, Sunny and Agarwal, Anshika and Agarwal, Harshvardhan and Aranguiz-Dias, Gabriela and Bhagirath, Aditri and Breuch, Justine and Chen, Huanxing and Chen, Ruishi and Chen, Sarah and Fan, Haocheng and Fang, William and Fergesen, Cat Gonzales and Frees, Daniel and Gao, Tian and Huang, Ziqing and Jain, Vishal and Jiang, Yucheng and Kalinin, Kirill and Karaca, Su Doga and Khatua, Arpandeep and La, Teland and Levent, Isabelle and Li, Miranda and Li, Xinling and Li, Yongce and Liu, Angela and Oh, Minsik and Paek, Nathan J. and Qin, Anthony and Redmond, Emily and Ryan, Michael J. and Salecha, Aadesh and Shen, Xiaoxian and Singhal, Pranava and Subrahmanya, Shashanka and Tan, Mei and Thawornbut, Irawadee and Vinocour, Michelle and Wang, Xiaoyue and Wang, Zheng and Weng, Henry Jin and Wirawarn, Pawan and Wu, Shirley and Wu, Sophie and Xie, Yichen and Ye, Patrick and Zhang, Sean and Zhang, Yutong and Zhou, Cathy and Zhao, Yiling and Landay, James and Yang, Diyi},
	year = {2026},
}

@article{sidelle_answering_1991,
	title = {The {Answering} {Machine} {Paradox}},
	volume = {21},
	number = {4},
	journal = {Canadian Journal of Philosophy},
	author = {Sidelle, Alan},
	year = {1991},
	pages = {525--539},
}

@article{cohen_indexicality_2013,
	title = {Indexicality and {The} {Answering} {Machine} {Paradox}},
	volume = {8},
	urldate = {2026-08-29},
	journal = {Philosophy Compass},
	author = {Cohen, Jonathan and Michaelson, Eliot},
	year = {2013},
	pages = {580--592},
}

@book{hofstadter_fluid_1996,
	title = {Fluid {Concepts} and {Creative} {Analogies}},
	publisher = {Basic Books},
	author = {Hofstadter, Douglas},
	year = {1996},
}

@article{eisenmann_machine_2024,
	title = {``{Machine} {Down}'': {Making} {Sense} of {Human--Computer Interaction}---{Garfinkel}'s {Research} on {ELIZA} and {LYRIC} from 1967 to 1969 and its {Contemporary Relevance}},
	volume = {39},
	number = {6},
	urldate = {2026-08-14},
	journal = {AI \& Society},
	author = {Eisenmann, Clemens and Mlyn\'{a}\v{r}, Jakub and Turowetz, Jason and Rawls, Anne W.},
	year = {2024},
	pages = {2715--2733},
}

@book{reeves_media_1996,
	address = {Cambridge},
	title = {The {Media} {Equation}: {How} {People} {Treat} {Computers}, {Television}, and {New} {Media} {Like} {Real} {People} and {Places}},
	publisher = {Cambridge University Press},
	author = {Reeves, Byron and Nass, Clifford},
	year = {1996},
}

@unpublished{radulescu_llms_nodate,
	title = {{LLMs} {Are} {Candidate} {Generators}},
	author = {Radulescu, Alexandru and Sterken, Rachel},
    note = {Manuscript},
    year = {2026}
}

@article{haslett_tokenization_2025,
	title = {Tokenization {Changes} {Meaning} in {Large} {Language} {Models}: {Evidence} from {Chinese}},
	volume = {51},
	number = {3},
	urldate = {2026-08-30},
	journal = {Computational Linguistics},
	author = {Haslett, David A.},
	year = {2025},
	pages = {785--814},
}

@misc{piantadosi_meaning_2022,
	title = {Meaning without reference in large language models},
	publisher = {arXiv},
	author = {Piantadosi, Steven T. and Hill, Felix},
	year = {2022},
}

@article{grindrod_large_2024,
	title = {Large {Language Models} and {Linguistic Intentionality}},
	volume = {204},
	number = {2},
	journal = {Synthese},
	author = {Grindrod, Jumbly},
	year = {2024},
	pages = {71},
}

@article{mandelkern_language_2024,
	title = {Do {Language} {Models}' {Words} {Refer}?},
	volume = {50},
	number = {3},
	journal = {Computational Linguistics},
	author = {Mandelkern, Matthew and Linzen, Tal},
	year = {2024},
	pages = {1191--1200},
}

@article{koch_babbling_2025,
	title = {Babbling stochastic parrots? {A} {Kripkean} argument for reference in large language models},
	volume = {1},
	journal = {Philosophy of AI},
	author = {Koch, Steffen},
	year = {2025},
	pages = {19--33},
}

@article{pepp_reference_2026,
	title = {Reference without intentions in large language models},
	volume = {69},
	number = {6},
	journal = {Inquiry},
	author = {Pepp, Jessica},
	year = {2026},
	pages = {2945--2963},
}

@inproceedings{bender_dangers_2021,
	address = {New York, N.Y.},
	series = {{FAccT} '21},
	title = {On the {Dangers} of {Stochastic} {Parrots}: {Can} {Language} {Models} {Be} {Too} {Big}? \emo{parrot}},
	booktitle = {Proceedings of the 2021 {ACM} {Conference} on {Fairness}, {Accountability}, and {Transparency}},
	publisher = {Association for Computing Machinery},
	author = {Bender, Emily M. and Gebru, Timnit and McMillan-Major, Angelina and Shmitchell, Shmargaret},
	year = {2021},
	pages = {610--623},
}

@article{titus_does_2024,
	title = {Does {ChatGPT} have semantic understanding? {A} problem with the statistics-of-occurrence strategy},
	volume = {83},
	journal = {Cognitive Systems Research},
	author = {Titus, Lisa Miracchi},
	year = {2024},
	pages = {101174},
}

@article{borg_llms_2026,
	title = {{LLMs}, {Turing} tests and {Chinese} rooms: the prospects for meaning in large language models},
	volume = {69},
	number = {6},
	journal = {Inquiry},
	author = {Borg, Emma},
	year = {2026},
	pages = {2807--2837},
}

@unpublished{yum_meaning_2026,
	title = {Meaning {Without} {Minds}: {A} {Conventionalist} {Account} of {Meaning} in {LLM} {Outputs}},
	author = {Yum, Inchul and Lee, Chanwoo and Park, Seong Soo},
	year = {2026},
	note = {Manuscript}
}

@incollection{kaplan_demonstratives_1989,
	title = {Demonstratives},
	booktitle = {Themes from {Kaplan}},
	publisher = {Oxford University Press},
    address = {Oxford},
	author = {Kaplan, David},
	editor = {Almog, Joseph and Perry, John and Wettstein, Howard},
	year = {1989},
	pages = {481--563},
}

@incollection{lewis_index_1981,
	address = {Dordrecht},
	title = {Index, {Context}, and {Content}},
	language = {en},
	booktitle = {Philosophy and {Grammar}: {Papers} on the {Occasion} of the {Quincentennial} of {Uppsala} {University}},
	publisher = {Springer Netherlands},
	author = {Lewis, Davis},
	editor = {Kanger, Stig and \"{O}hman, Sven},
	year = {1981},
	pages = {79--100},
}

@incollection{braun_indexicals_2026,
	edition = {{Fall} 2026},
	title = {Indexicals},
	url = {https://plato.stanford.edu/archives/fall2026/entries/indexicals/},
	booktitle = {The {Stanford} {Encyclopedia} of {Philosophy}},
	publisher = {Stanford University},
	author = {Braun, David and Radulescu, Alexandru},
	editor = {Zalta, Edward N. and Nodelman, Uri},
	year = {2026},
}

@incollection{kaplan_afterthoughts_1989,
	title = {Afterthoughts},
	booktitle = {Themes from {Kaplan}},
	publisher = {Oxford University Press},
    address = {Oxford},
	author = {Kaplan, David},
	editor = {Almog, Joseph and Perry, John and Wettstein, Howard},
	year = {1989},
	pages = {565--614},
}

@unpublished{cappelen_room_nodate,
	title = {The {Room} of {Amnesiacs}: {AI} {Individuation} and {Reference}},
	shorttitle = {The {Room} of {Amnesiacs}},
	author = {Cappelen, Herman and Hawthorne, John},
    journal = {Inquiry},
    note = {Manuscript},
    year = {2026}
}

@book{stojnic_context_2021,
	address = {Oxford, New York},
	title = {Context and {Coherence}},
	publisher = {Oxford University Press},
	author = {Stojni\'{c}, Una},
	year = {2021},
}

@article{cappelen_creative_2008,
	title = {The {Creative} {Interpreter}: {Content} {Relativism} and {Assertion}},
	volume = {22},
	journal = {Philosophical Perspectives},
	author = {Cappelen, Herman},
	year = {2008},
	pages = {23--46},
}

@article{egan_billboards_2009,
	title = {Billboards, bombs and shotgun weddings},
	volume = {166},
	number = {2},
	journal = {Synthese},
	author = {Egan, Andy},
	year = {2009},
	pages = {251--279},
}

@book{cappelen_making_2021,
	title = {Making {AI} {Intelligible}: {Philosophical} {Foundations}},
	publisher = {Oxford University Press},
    address = {Oxford},
	author = {Cappelen, Herman and Dever, Josh},
	year = {2021},
}

@incollection{king_content_2005,
    author = {King, Jeffrey C. and Jason Stanley},
    title = {Semantics, Pragmatics, and the Role of Semantic Content},
    booktitle = {Semantics Versus Pragmatics},
    publisher = {Oxford University Press},
    address = {Oxford},
    year = {2005},
    editor = {Szab\'{o}, Zolt\'{a}n Gendler},
    pages = {111--164}
}

@book{austin_how_1962,
	address = {Oxford},
	title = {How to {Do} {Things} with {Words}},
	publisher = {Oxford University Press},
	author = {Austin, J.L.},
	editor = {{Sbisa, Marina}},
	year = {1962},
}

@article{lewis_scorekeeping_1979,
	title = {Scorekeeping in a language game},
	volume = {8},
	number = {1},
	journal = {Journal of Philosophical Logic},
	author = {Lewis, David},
	year = {1979},
	pages = {339--359},
}

@article{stalnaker_common_2002,
	title = {Common {Ground}},
	volume = {25},
	number = {5/6},
	journal = {Linguistics and Philosophy},
	author = {Stalnaker, Robert},
	year = {2002},
	pages = {701--721},
}

@article{roberts_information_2012,
	title = {Information {Structure}: {Towards} an integrated formal theory of pragmatics},
	volume = {5},
	journal = {Semantics and Pragmatics},
	author = {Roberts, Craige},
	year = {2012},
	pages = {1--69},
}

@incollection{resnick_grounding_1991,
	address = {Washington},
	title = {Grounding in {Communication}},
	booktitle = {Perspectives on {Socially Shared Cognition}},
	publisher = {American Psychological Association},
	author = {Clark, Herbert H. and Brennan, Susan E.},
	editor = {Resnick, Lauren B. and Levine, John M. and Teasley, Stephanie D.},
	year = {1991},
	pages = {127--149},
}

@unpublished{almotahari_cooperative_2025,
	title = {Cooperative {Speech}, {Semantic} {Competence}, and {AI}},
	url = {https://arxiv.org/abs/2504.16092v1},
	author = {Almotahari, Mahrad},
	year = {2025},
    note = {Manuscript}
}

@incollection{grice_logic_1975,
	address = {New York, N.Y.},
	title = {Logic and {Conversation}},
	volume = {3},
	booktitle = {Syntax and {Semantics}},
	publisher = {Academic Press},
	author = {Grice, H. P.},
	editor = {Cole, Peter and Morgan, Jerry L},
	year = {1975},
	pages = {1--58},
}

@book{quine_word_1960,
	title = {Word and {Object}},
	publisher = {The MIT Press},
    address = {Cambridge, MA.},
	author = {Quine, Willard Van Orman},
	year = {1960},
}

@article{davidson_radical_1973,
	title = {Radical {Interpretation}},
	volume = {27},
	number = {3/4},
	journal = {Dialectica},
	author = {Davidson, Donald},
	year = {1973},
	pages = {313--328},
}

@article{mallory_fictionalism_2023,
	title = {Fictionalism about {Chatbots}},
	volume = {10},
    number = {38},
	journal = {Ergo an Open Access Journal of Philosophy},
	author = {Mallory, Fintan},
	year = {2023},
}

@article{ruyant_what_2026,
	title = {What do large language models represent?},
	volume = {207},
	journal = {Synthese},
	author = {Ruyant, Quentin},
	year = {2026},
	pages = {17},
}

@incollection{proudfoot_communicating_2026,
	title = {Communicating with {AI}? {Wittgenstein}'s {Lion} and the {Problem} of {Cross}-{Species} {Communication}},
	booktitle = {Communicating with {AI}: {Philosophical} {Perspectives}},
	publisher = {Oxford University Press},
    address = {Oxford},
	author = {Proudfoot, Diane},
	editor = {Cappelen, Herman and Sterken, Rachel},
	year = {2026},
	pages = {194--219},
}

@misc{marks_persona_2026,
	title = {The {Persona} {Selection} {Model}: {Why} {AI} {Assistants} might {Behave} like {Humans}},
	url = {https://alignment.anthropic.com/2026/psm/},
	urldate = {2026-09-03},
	journal = {Alignment Science Blog},
	author = {Marks, Sam and Lindsey, Jack and Olah, Christopher},
	year = {2026},
}

\end{document}